\documentclass[
 reprint,
 amsmath,amssymb,
 aps,
]{revtex4-2}

\usepackage{graphicx}
\usepackage{dcolumn}
\usepackage{bm}
\usepackage{hyperref}

\usepackage{url}

\begin{document}

\preprint{APS/123-QED}

\title{Interplay of epidemic spreading and vaccine uptake under complex social contagion}

\author{Alfonso de Miguel-Arribas}
\email{alfonso.demiguel.arribas@gmail.com}
\affiliation{
 Institute for Biocomputation and Physics of Complex Systems (BIFI), University of Zaragoza, 50018, Zaragoza, Spain.
}
\affiliation{
 Zaragoza Logistics Center (ZLC), Av. de Ranillas 5, 50018, Zaragoza, Spain
}

\author{Alberto Aleta}
\affiliation{
 Institute for Biocomputation and Physics of Complex Systems (BIFI), University of Zaragoza, 50018, Zaragoza, Spain.
}
\affiliation{
 Department of Theoretical Physics, University of Zaragoza, 50018, Zaragoza, Spain.
}

\author{Yamir Moreno}
\affiliation{
 Institute for Biocomputation and Physics of Complex Systems (BIFI), University of Zaragoza, 50018, Zaragoza, Spain.
}
\affiliation{
 Department of Theoretical Physics, University of Zaragoza, 50018, Zaragoza, Spain.
}
\affiliation{
 Centai Institute, Turin, Italy
}

\date{\today}

\begin{abstract}
Modeling human behavior is essential to accurately predict epidemic spread, with behaviors like vaccine hesitancy complicating control efforts. While epidemic spread is often treated as simple contagion, vaccine uptake may follow complex contagion dynamics, where individuals’ decisions depend on multiple social contacts. Recently, the concept of complex contagion has received strong theoretical underpinnings thanks to the generalization of spreading phenomena from pairwise to higher-order interactions. Although several potential applications have been suggested, examples of complex contagions motivated by real data remain scarce. Surveys on COVID-19 vaccine hesitancy in the US suggest that vaccination attitudes may indeed depend on the vaccination status of social peers, aligning with complex contagion principles. In this work, we examine the interactions between epidemic spread, vaccination, and vaccine uptake attitudes under complex contagion. Using the SIR model with a dynamic, threshold-based vaccination campaign, we simulate scenarios on an age-structured multilayer network informed by US contact data. Our results offer insights into the role of social dynamics in shaping vaccination behavior and epidemic outcomes.
\end{abstract}


\maketitle

\section{Introduction}
\label{sec:intro}

The integration of human behavior into the modeling of epidemic spreading has long drawn researchers' attention \cite{ferguson2007capturing}. Human behavior fundamentally influences the spread of epidemics in various ways. These include how our societies, mindsets, and lifestyles are organized, how human settlements tend to favor crowding (both human and animal), as well as our modes of transportation, which facilitate the fast and global spread of otherwise local diseases. Additionally, our responses to epidemic outbreaks play an important role. Particularly in this latter aspect, modeling human behavior remains a formidable challenge due to its complex nature and the lack of real-world, quantitative data on behavioral changes in populations affected by epidemic outbreaks \cite{funk2010modelling, perra2011towards}.

Key aspects of human behavior that significantly shape epidemics are the fear of infection, information awareness, and behavioral attitudes towards control interventions \cite{vardavas2007can, epstein2008coupled, salathe2008effect, perra2011towards, mao2012coupling, rizzo2014effect}. In particular, vaccination provokes hesitant attitudes in some sectors of the population, diminishing its efficacy at the population level \cite{dube2013vaccine, troiano2021vaccine}. For this reason, the dynamics of vaccination has long been a topic within the epidemic modeling literature \cite{bernoulli1760essai, hethcote1978immunization, liu2008svir, hsieh2010age, wang2016statistical}, and numerous works have attempted to incorporate genuine behavioral features into the vaccination process \cite{vardavas2007can, mao2012coupling, wang2015coupled, wang2016statistical, yin2022impact}.
 
Despite the lack of a well-established unifying framework to couple epidemic spreading and social dynamics, one of the standard approaches to integrate human behavior into epidemiological models is game theory through the so-called vaccination games \cite{chang2020game, tanimoto2021sociophysics}. Here, individuals decide whether to vaccinate or not depending on the assessment of the cost incurred by the options they face. This assessment although traditionally assumed to be perfectly rational and mostly isolated from the social environment, has been progressively extended to incorporate more realistic features that affect the human decision-making process \cite{wang2015coupled, iwamura2018realistic, chang2020game, tanimoto2021sociophysics}. Another approach for integrating epidemic models and vaccination-related behavior, significantly less exploited than vaccination games, is that of opinion dynamics \cite{castellano2009statistical}. In this approach, individuals hold an opinion on a topic (such as vaccine uptake), which can change or be influenced through various mechanisms, including peer interaction and social and mass media exposure. Notable references include \cite{ni2011modeling, alvarez2017epidemic, pires2017dynamics, pires2018sudden, pires2021antivax, fugenschuh2022overcoming, teslya2022effect}, where complex behaviors involving sudden transitions, bistability, or network segregation have been demonstrated.

Among the most notable models of opinion dynamics, lies the seminal Watts-Granovetter threshold model \cite{granovetter1978threshold, watts2002simple, watts2007influentials}. Originally aimed at modeling phenomena such as riots and other social movements, this model was later generalized to the study of information cascades, stock market crashes, and cascading failures in infrastructure networks. These collective phenomena are typically referred to as \textit{complex contagions} \cite{centola2007complex}. Unlike simple contagions, where the transmission of infection occurs at a rate between individuals and a single infected neighbor is always sufficient to expose a susceptible node, in complex contagions, the behavior of an individual is conditional on the behavior of a fraction of their peers \cite{centola2007complex, min2018competing, cencetti2023distinguishing}. Thus, while biological processes such as epidemics may be seen as simple contagion processes, human behavior dynamics related to opinions and decisions on matters of self-protection and vaccine uptake might constitute complex contagions. Recent studies have indeed begun to explore the interaction of simple and complex contagions within the context of epidemics \cite{salathe2008effect, salathe2011assessing, campbell2013complex, fu2017dueling, centola2020complex, hebert2020spread, fugenschuh2022overcoming, qiu2022understanding}.

At the same time, a substantial body of theoretical work is being developed around the concept of complex contagion, through the generalization of spreading phenomena from pairwise to higher-order interactions, extending traditional networks to structures such as simplicial complexes and hypergraphs \cite{iacopini2019simplicial, de2020social, battiston2021physics}. Currently, though several potential applications have been suggested \cite{ferraz2024contagion}, examples motivated by real data where complex contagion is at play are scarce. In this regard, we have found explicit evidence from COVID-19 vaccine hesitancy surveys \cite{lazer2021covid}, suggesting that people's attitudes towards vaccine uptake could depend on the behavior of their social peers, perfectly aligning with the concept of a complex contagion process.

Motivated by these observations, in this work, we investigate the interplay between the spread of an epidemic on networks and the dynamics of opinion on vaccine uptake following a complex contagion interaction. We employ the standard SIR model, which we couple with a dynamic vaccination campaign. The progression of this campaign is, in turn, influenced by a threshold-based opinion dynamics process, effectively integrating models of simple and complex contagion. To offer a more realistic setting where the dynamics unfold, we use an age-based multilayer contact network substrate, generated from data-driven contact matrices in the United States \cite{mistry2021inferring}. We investigate the system behavior by first assuming that everyone has the same activation threshold to switch their opinion towards a pro-vaccine stance, and later we explore the heterogeneous landscape of vaccine attitudes across the US as provided by Lazer et al. surveys \cite{lazer2021covid}.

\section{Model}
\label{sec:model}

\subsection{Definition}
\label{subsec:definition}

Our model consists of coupled dynamics featuring a simple epidemic process, a vaccination campaign, and opinion dynamics based on a complex contagion model, running on an age-layered contact network. 

This multilayer network is generated from contact matrices constructed from age-based data-driven synthetic populations \cite{mistry2021inferring, aleta2020data}. Thus, the nodes in layer $a$ represent individuals of age $a$, whose contacts are distributed both in the same age layer and across the rest of them, following realistic contact patterns. Our geographical focus here is on the United States, both at the national level and on a state basis.

As a simple approximation to influenza-like illnesses, we take the conventional susceptible-infected-removed SIR model for the epidemic process:
\begin{equation}
\mathcal{S}+\mathcal{I}\xrightarrow{\beta}\mathcal{I}+\mathcal{I},
\end{equation}
\begin{equation}
\mathcal{I}\xrightarrow{\mu}\mathcal{R},
\end{equation}
where $\beta$ is the disease's transmission rate, and $\mu$ is the recovery rate. Throughout this work, we fix the basic reproduction number at $R_0=1.5$, representing a moderate transmissibility scenario, and set $\mu=0.2$. The transmission rate $\beta$ is then adjusted following the relationship $R_0 \approx (\beta/\mu)\langle k\rangle$, where $\langle k\rangle$ denotes the average degree of the contact structure, which is approximately $\langle k\rangle = 12$ across all the states in the country.

All susceptible individuals are eligible to an ongoing vaccination campaign, modeled also as a Poisson process with rate $\alpha$:
\begin{equation}
\mathcal{S}\xrightarrow{\alpha}\mathcal{V}.
\end{equation}
This parameter $\alpha$ can be interpreted as the vaccination effort carried out by the public health authorities during the vaccination campaign.
However, not all susceptible individuals will automatically proceed to be vaccinated, since this is conditioned to the individual's attitude towards this control measure.

Inspired by surveys on vaccine hesitancy conducted at the beginning of the COVID-19 vaccination campaign in the US, we add a third dynamical process in the form of opinion dynamics under complex contagion. The attitude towards vaccination is binary: inactive (hesitant) $\mathcal{H}$ or (pro-)active $\mathcal{A}$. The following transition is defined:
\begin{equation}
\mathcal{H}\xrightarrow{}\mathcal{A}.
\end{equation}
Thus, hesitant behavior is not necessarily anti-vaccine behavior and once an individual has decided to be vaccinated, they will not change their mind. The specific mechanism under which this change of opinion proceeds is through a threshold-based updating rule in the vein of the renowned Watts-Granovetter model \cite{watts2007influentials}:
\begin{equation}
o_i(t+1)=
    \begin{cases}
        1 & \text{if } \frac{V_{\Omega_i}(t)}{k_i}\geq\theta_i,\\
        0 & \text{otherwise}.
    \end{cases}
\end{equation}
Here $\theta_i$ is the individual's $i$ activation threshold, $k_i$ is the number of first-neighbors of $i$, and $V_{\Omega_i}(t)$ is the total number of vaccinated individuals within these neighbors. This relationship directly connects the success of the ongoing vaccination campaign to the opinion on vaccination, setting a co-evolving feedback loop between vaccination and opinion dynamics. 

It must be noted that no age-dependence has been made explicit so far since all these processes and parameters are equal for any individual in the system. In our model, age enters implicitly through the specific contact patterns of the multilayer network. The reason for this is to focus directly on the global impact of the coupled simple and complex contagion dynamics, without confounding features.

\begin{figure*}
\centering
\includegraphics[width=\textwidth]{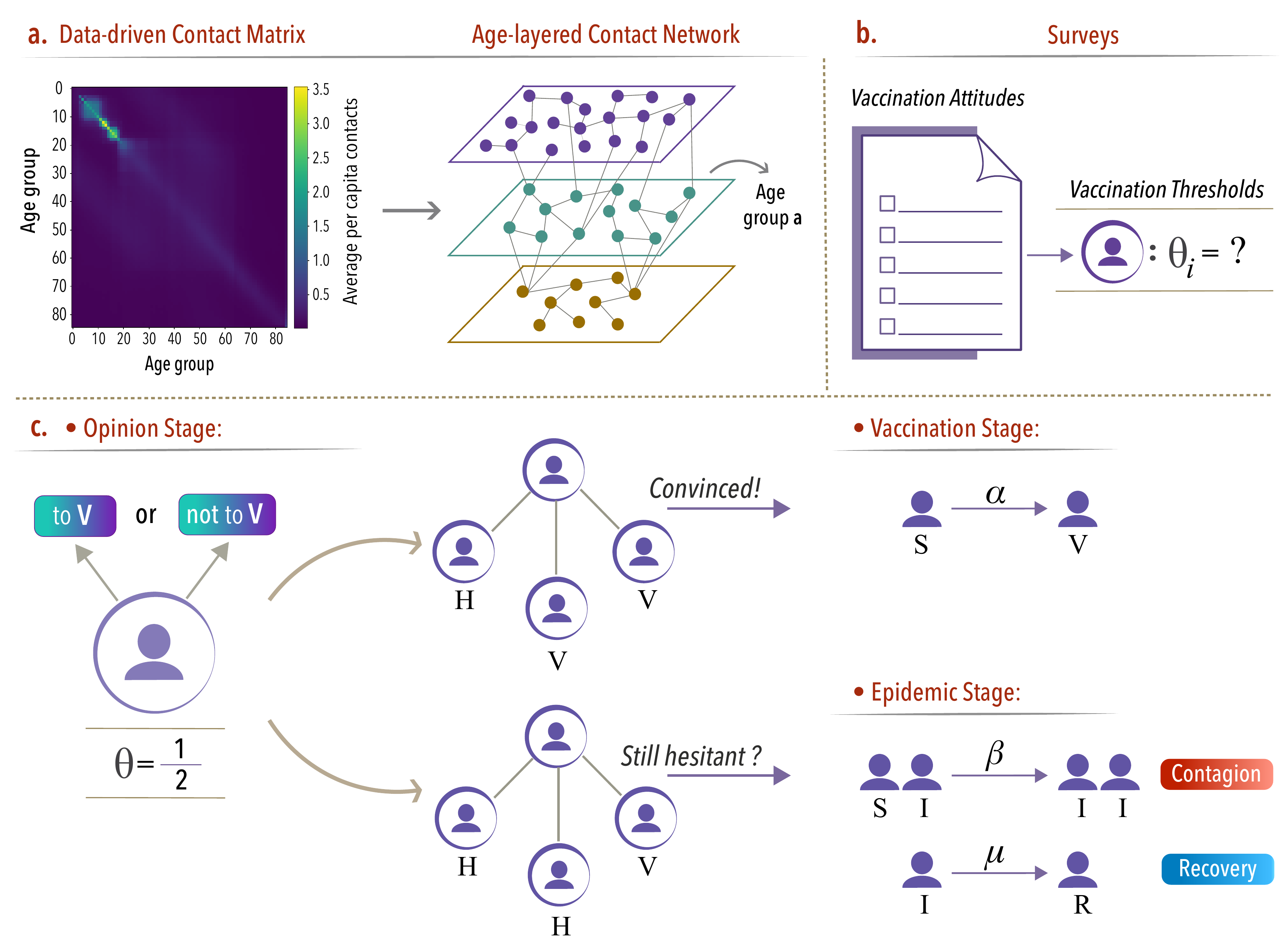}
\caption{\label{fig:main_approach}
\textbf{Modeling approach scheme}. (a) Contact matrices are constructed using data-driven synthetic populations \cite{mistry2021inferring} and are used to generate age-layered contact networks following the approach proposed in \cite{aleta2020data}. (b) Vaccination surveys conducted during the COVID-19 pandemic by Lazer et al. \cite{lazer2021covid} inform the conversion of vaccination attitudes into numerical vaccination thresholds. Initially, individuals are assumed to have homogeneous vaccination thresholds. In a later stage of the work, these thresholds are refined using the data-informed values from the surveys. (c) With the contact structure established and vaccination thresholds assigned to individuals, the dynamics of the model are initiated. During the opinion stage, each individual decides whether to be vaccinated based on their threshold and the vaccination status of their neighbors. If the proportion of vaccinated neighbors exceeds the threshold $\theta$, the individual becomes vaccinated with a probability $\alpha$. Otherwise, they remain unvaccinated and participate in the epidemic stage, governed by a standard SIR model.}
\end{figure*}

Our approach is summarized in Figure \ref{fig:main_approach}. More details on the specifics of the data and modeling approach can be found in the Methods, section \ref{sec:methods}.

\subsection{Experiments}
\label{subsec:experiments}

We run extensive discrete-time Monte Carlo simulations of the coupled dynamics on the multilayer structure.

The system behavior is characterized by these main observables: the number of active (convinced to vaccinate) agents $N_A(\infty)$, the prevalence $R(\infty)$, that is, the fraction of the population affected by the disease, and the vaccination coverage (VC) $V(\infty)$, the number of people that end up taking the vaccine. All these quantities are referred to the equilibrium state, that is when the fraction of infected individuals in the system reaches zero. By normalizing these quantities by the system's size $N$, we represent them as $r(\infty)=R(\infty)/N$, $v(\infty)=V(\infty)/N$, and $n_A(\infty)=N_A(\infty)/N$, respectively. These quantities will be examined across varying control parameters such as the activation threshold $\theta$, the initial fraction of active individuals $n_A(0)$, and the vaccination rate $\alpha$. As a nuance, given that the outlook of $n_A(\infty)$ and $v(\infty)$ is very similar, and the fact that $n_A(\infty)$ is highly conditioned on $n_A(0)$, we define the following derived quantity:
\begin{equation}
    \Delta n_A(\infty)\equiv\frac{n_A(\infty)-n_A(0)}{1-n_A(0)},
\end{equation}
which can be interpreted as a measure of the normalized final size of the behavioral cascade triggered given the initial condition $n_A(0)$. In other words, $\Delta n_A(\infty)$ informs us about the change in vaccine support relative to the initial support $n_A(0)$. It must be noted that the distribution of this variable $n_A(0)$ is assigned uniformly at random, that is, the initial set of individuals with a pro-active attitude towards vaccine uptake is independent of their age, degree, or opinion assortativity. 

Finally, we consider vaccination rates ranging from $\alpha=0.001$, which amounts to vaccinating $0.1\%$ of the eligible population, to $\alpha=0.05$. To put these numbers in perspective, the highest daily vaccination record in the United States, according to \cite{owid_vaccine_rates}, was $3.15$ million doses on April 11, 2021, when the total population that had received two doses was approximately $73$ million, as reported by \cite{us_facts_vaccines}. Given that the total US population is about $350$ million, this represents a rate of roughly $0.012$ on that record day ($1.2\%$ of the population). Nonetheless, actual vaccination rates are much more variable as compared to our simplified model of a constant rate. From a global perspective, during the COVID-19 pandemic, despite the urgent vaccination deployment efforts and increased disease awareness, the daily percentage of the population receiving a COVID-19 vaccine dose rarely exceeded $1\%$, equivalent to $\alpha=0.01$, with occasional exceptions in countries like China and the UK.  As reported in \cite{owid_vaccine_rates2}, these countries experienced peaks of $1.57\%$ and $1.41\%$ of their populations getting vaccinated in a single day, respectively. These figures are instrumental in contextualizing our vaccination rate scenarios.

\section{Results}
\label{sec:results}

\subsection{Homogeneous thresholds}
\label{subsec:homogeneous}

\begin{figure*}
\centering
\includegraphics[width=\textwidth]{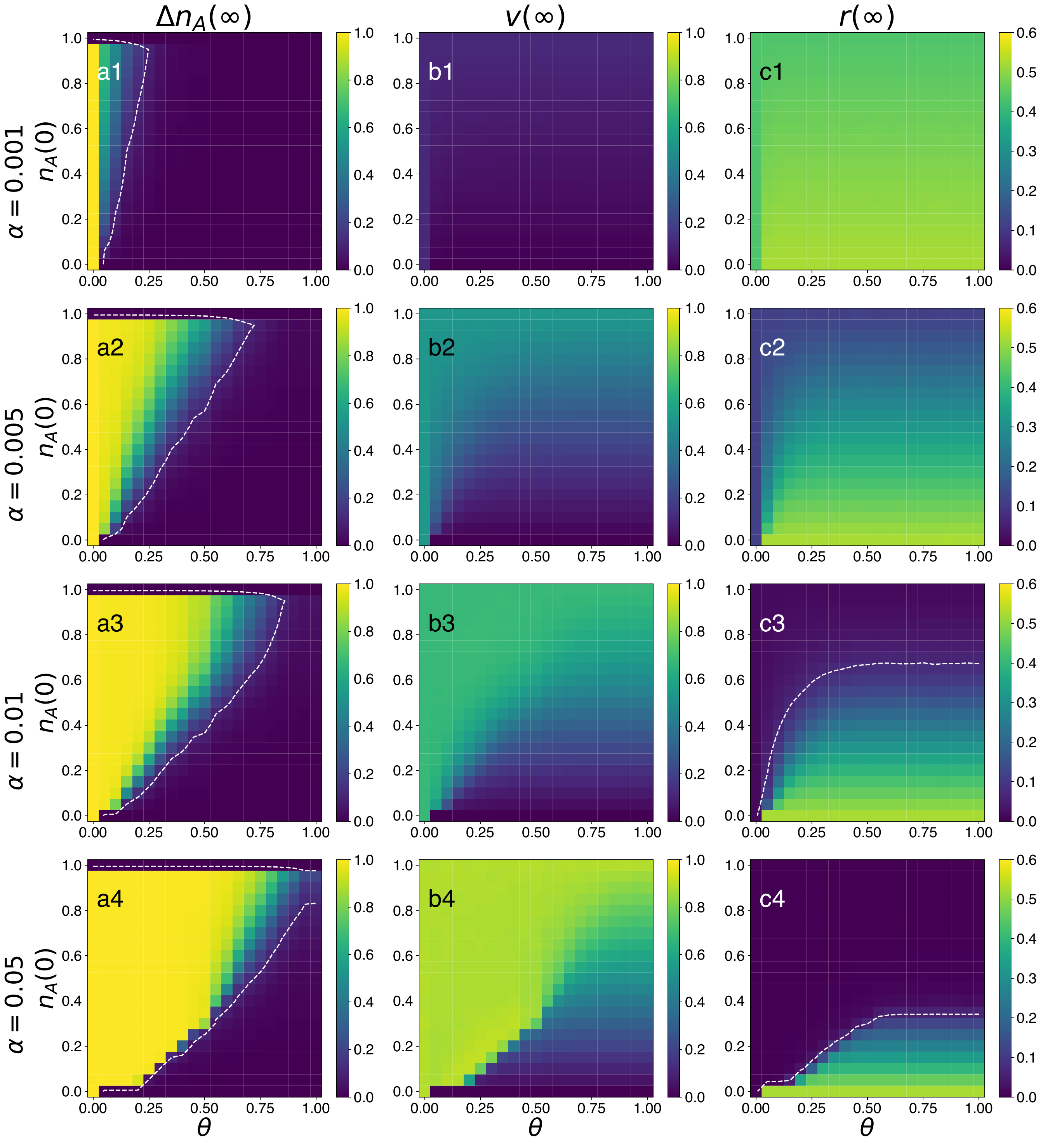}
\caption[Homogeneous thresholds for the US multilayer network]{\label{fig:main_homogeneous_heatmap} \textbf{System's behavior under homogeneous thresholds for the US multilayer network.}  Results for normalized activation change $\Delta n_A(\infty)$ (Column \textbf{a}), vaccination coverage $v(\infty)$ (Column \textbf{b}), and prevalence $r(\infty)$ (Column \textbf{c}). In every panel, outcomes are explored in $(\theta,n_A(0))$ space. Every row from 1 to 4 shows the solution for a different vaccination scenario as defined by the vaccination rate $\alpha$. Every point in the diagrams amounts to $25$ network realizations times $50$ dynamical realizations. White dashed line contours mark an approximate (due to finite size effects) separation between $\Delta n_A(\infty)=0$ or $r(\infty)=0$ and non-null values of the respective observables.}
\end{figure*}

Initially, we study the system's behavior under a homogeneous threshold $\theta$ scenario (Figure \ref{fig:main_homogeneous_heatmap}). The parameter space related to the opinion dynamics process, $(\theta, n_A(0))$ is scrutinized in finer detail, and we select certain vaccination scenarios based on the rate $\alpha$. As representative enough, the US contact patterns data at the national level are utilized to generate the underlying multilayer contact network.

Four different vaccination scenarios with substantially different vaccination efforts are considered, ranging from $\alpha=0.001$ (Panels a1-c1), $\alpha=0.005$ (Panels a2-c2), $\alpha=0.01$ (Panels a3-c3), to $\alpha=0.05$ (Panels a4-c4). As framed by the prior discussion on COVID-19 vaccination rates, the scenarios proposed here lie from a very low effort, $\alpha=0.001$, to a very high and rather unrealistic vaccine unroll, $\alpha=0.05$, so that we can have insights on the system behavior under extreme but meaningful vaccination bounds. 

For $\alpha=0.001$, we can appreciate how $\Delta n_A(\infty)\approx 0$ for the most part of $(\theta,n_A(0))$-space (Figure \ref{fig:main_homogeneous_heatmap}a1), this meaning that $\alpha$ is very low to trigger a substantial vaccination coverage (Figure \ref{fig:main_homogeneous_heatmap}b1), remaining very close to null, and thus system's prevalence tends to be rather homogeneous and maximal across the parameter space (Figure \ref{fig:main_homogeneous_heatmap}c1). The poor vaccination coverage cannot bring down the system's endemic phase in any point of the parameter space. Notably increasing $\alpha$ to $\alpha=0.005$, a different picture starts to emerge. The region where $\Delta n_A(\infty)\neq 0$ extends toward higher $n_A(0)$ and $\theta$ (Figure \ref{fig:main_homogeneous_heatmap}a2). These behavioral changes lead to mass vaccine uptake when initial vaccine support $n_A(0)$ is high enough, independently of the activation threshold $\theta$ (Figure \ref{fig:main_homogeneous_heatmap}b2). Consequently, the disease's prevalence decreases in that same region (Figure \ref{fig:main_homogeneous_heatmap}c2). Doubling $\alpha$ to reach $\alpha=0.01$, leads to a qualitatively different picture when looking at the system's prevalence (Figure \ref{fig:main_homogeneous_heatmap}c3). Vaccination coverage $v(\infty)$ is high enough in an important part of the system that yields a disease-free phase for medium-high $n_A(0)$ and also for lower values of this initial support as long as $\theta$ is kept very small. Finally, in the very high effort scenario, $\alpha=0.05$, the trend continues, as the disease-free phase extends to lower values of $n_A(0)$ and higher values of $\theta$. 

Overall, for a fixed value of $\theta$, moving from $n_A(0)=0$ to $n_A(0)=1$, means that vaccine hesitancy loses ground, VC increases and disease prevalence tends to zero. In high enough $\alpha$ scenarios, for low $\theta$, the transitions occur in a more abrupt way, whereas for medium to high $\theta$, the transitions are smoother. Varying $\theta$ with fixed $n_A(0)$ does not have a notable effect except where a phase separation exists. If the initial vaccine acceptance $n_A(0)$ is high enough, even $\theta\to 1$ has no effect on vaccination and prevalence. A giant component of vaccinated agents can emerge fast enough to avoid a sizeable outbreak. On the other hand, as $n_A(0)$ decreases (for a fixed $\alpha$), critical values of $\theta$ appear that, if surpassed, can bring the system from a disease-free phase to an endemic phase. This critical threshold $\theta$, however, can be pushed toward higher values if the vaccination campaign proceeds at faster rates. 

Importantly, though, even in the most intensive vaccination scenario explored, the epidemiological impact could be sizeable if the initial support for vaccination is rather mild, and the threshold to be convinced is not very high. This is clearly a hindrance product of social contagion being regarded as a complex contagion process rather than a simple one, requiring greater support to diffuse a pro-active attitude towards vaccine uptake.

Complementarily, in the appendix \ref{app:sections}, we offer curves of the explored observables for varying $\theta$ with sections of constant $n_A(0)$, where the abrupt nature of the changes in $\Delta n_A(\infty)$, $v(\infty)$, and $r(\infty)$ is better appreciated. 

\subsection{Heterogeneous thresholds from surveys}
\label{subsec:heterogeneous}

In the second stage of this work, we inform our model with activation thresholds derived from real data collected from the surveys carried out by Lazer et al. \cite{lazer2021covid}, at the beginning of the COVID-19 vaccination campaign in the US. These surveys offer, among other features, insights into vaccination acceptance and hesitancy, segmented by age at the national level, and also account for the fraction of people within a certain attitude towards vaccine uptake at the state level.

The attitudes are categorical and include: ``already vaccinated", inclined to be vaccinated ``as soon as possible", ``after at least some people I know", ``after most people I know", and those who ``would not get the COVID-19 vaccine". It is from this threshold-based perspective and its heterogeneous distribution—varying according to others' vaccination statuses—that we draw inspiration for our coupled epidemic-vaccination-opinion dynamics model.

Now, since the control parameters $\theta$ and $n_A(0)$ previously used are now fixed for every state, the only free parameter is the vaccination rate $\alpha$, which we use to extensively explore the outcomes of the coupled dynamics. Taking full advantage of the vaccination attitude data for every state in the US, simulations are conducted for every US state. It is important to note that the assignment of different vaccination attitudes and their corresponding thresholds to agents in the network is done randomly.

\subsubsection{System's behavior under different vaccination scenarios}

\begin{figure*}
\centering
\includegraphics[width=\textwidth]{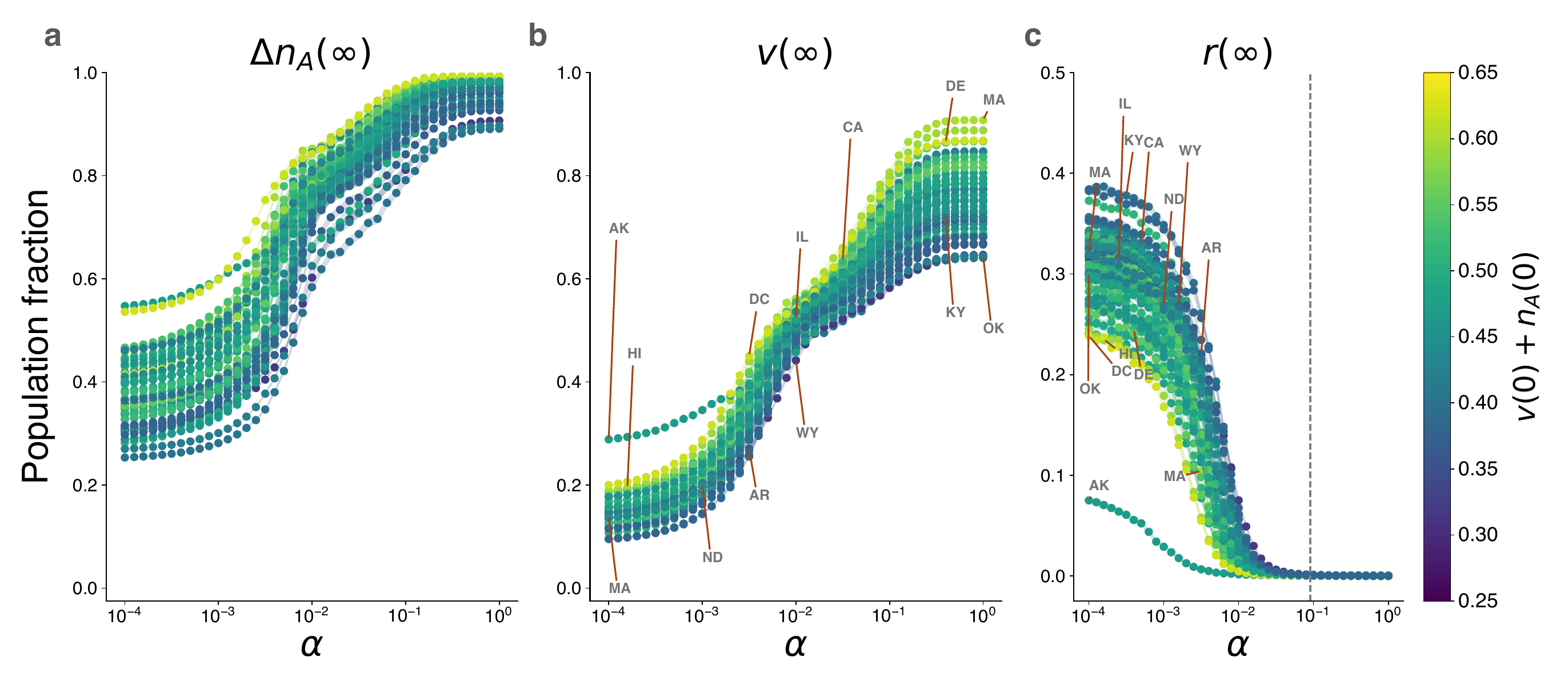}
\caption{\label{fig:main_vaccination_curves} \textbf{Epidemiological impact across US states for varying vaccination rates $\alpha$}. The vertical dashed line represents an approximate separation between the endemic phase and the disease-free phase. Panel a: $\Delta n_A(\infty)$ vs. $\alpha$. Panel b: $v(\infty)$ vs. $\alpha$. Panel c: $r(\infty)$ vs. $\alpha$. The color bar represents the initial fraction of support for the vaccine: ``already vaccinated'' $v(0)$ plus ``as soon as possible'' $n_A(0)$. Some state labels are shown for reference.}
\end{figure*}

Similarly as before, we examine how $\Delta n_A(\infty)$, $v(\infty)$, and $r(\infty)$ behave across varying $\alpha$ values in Figure \ref{fig:main_vaccination_curves}, panels a, b, and c, respectively. 

We observe that $\Delta n_A(\infty)$ across states remains between $0.2$ and $0.6$ for the vaccination scenarios with the lowest effort. This separation diminishes as $\alpha$ increases, aligning within the range of $0.9$ to $1.0$ for scenarios with the highest vaccination rate $\alpha$. Notably, no state exhibits $\Delta n_A(\infty)=0$ in any scenario, implying, that all of them can trigger behavioral cascades that increase the number of convinced individuals and thus there is substantial potential for vaccination coverage to grow. Regarding vaccination coverage, we appreciate how in low $\alpha$ regimes, VC is primarily between $0.1$ and $0.2$, with the exception of Alaska (AK). For intermediate $\alpha$ values, representing more realistic vaccine uptake rates, the disparity in VC across states reduces but widens again as $\alpha\to 1$. Due to inherent differences in each state's contact network but, more importantly, their $\theta$ distribution, the curves show different rates of growth and intersect at various vaccination scenarios. This phenomenon reflects the diversity in vaccine attitudes across states. As a specific example, the case of Massachusetts (MA) is especially interesting. In the lowest $\alpha$ regimes, MA occupies an average rank among states, with $13\%$ of its population having already received the vaccine. However, with increasing $\alpha$, the vaccination-opinion feedback loop is further stimulated. This facilitates the progressive vaccination of individuals in subsequent categories, specifically ``soon'' at $47\%$, and ``someone'' at $18\%$. Ultimately, this leads MA to emerge as one of the states with the highest vaccination coverage. The disease's prevalence, consequently, is the highest for the lowest vaccination rates, spanning approximately from $0.25$ to $0.4$. These values rapidly decrease with increasing vaccine unroll, reaching a disease-free state beyond $\alpha\approx 0.05$. Alaska is correspondingly an exception here too; its prevalence is an outlier for the lowest $\alpha$ scenarios ($r(\infty)<0.1$), and sizeable outbreaks are negligible for $\alpha>0.005$. With an initial ``already vaccinated'' fraction of $v(0)=0.29$, Alaska accounts for the highest initial immunized population. 

All the curves across the panels incorporate additional information represented by a color map for the sum of initial conditions for vaccinated and pro-vaccine individuals, $v(0)+n_A(0)$. Overall, there is a positive correlation between this initial condition and VC, and a negative one towards the disease's prevalence.

\subsubsection{Relevance of heterogeneous thresholds}

To explore the effect of heterogeneity in vaccination attitudes, let us examine more closely the relevance of each vaccination attitude, and thus the threshold $\theta$, in determining the final epidemic size $r(\infty)$.

\begin{figure*}
\centering
\includegraphics[width=\textwidth]{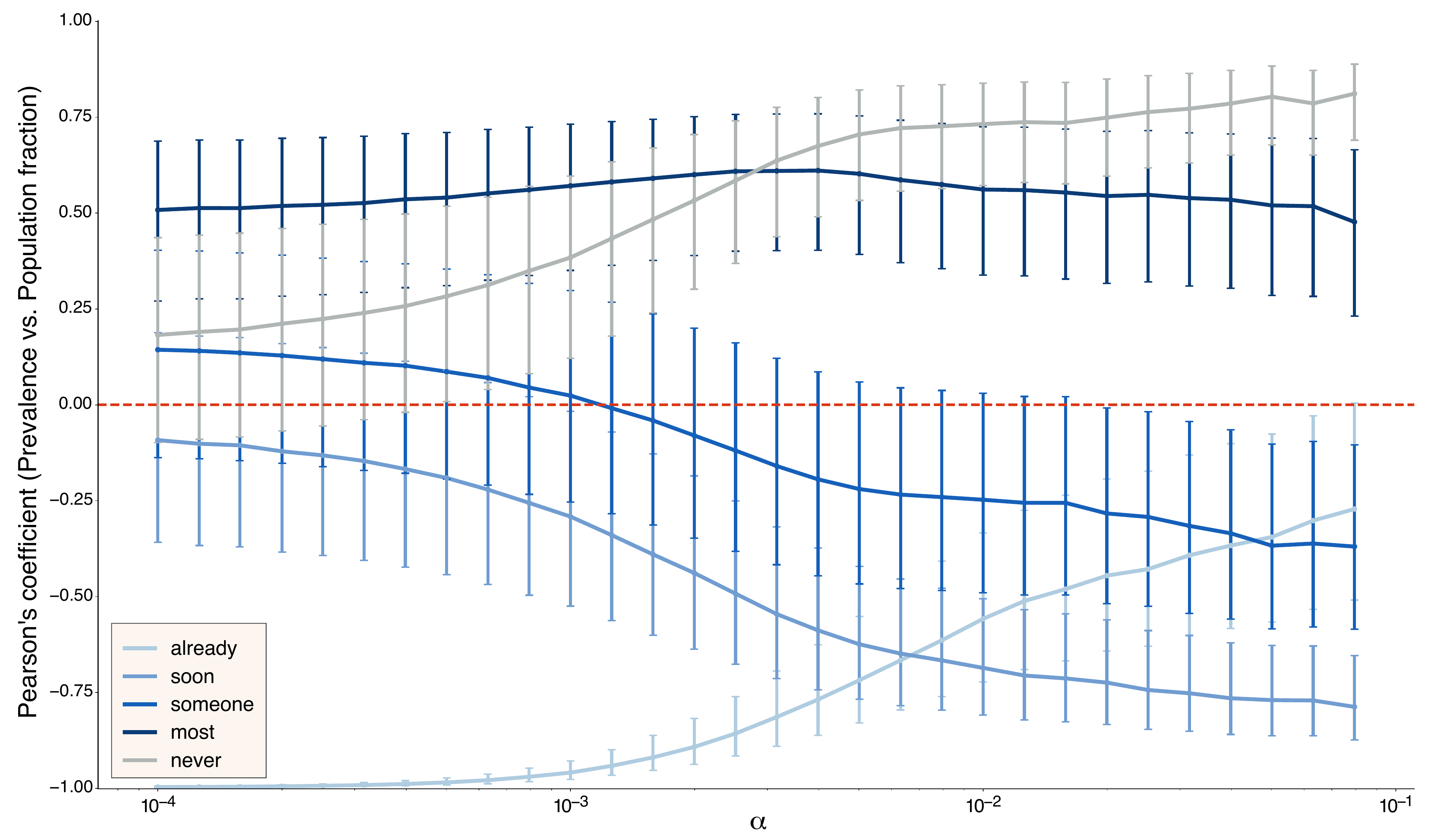}
\caption{\label{fig:main_correlation_curves} \textbf{Pearson's coefficient (Prevalence vs. Population fraction) as a function of vaccination rates}. The coefficient is measured using a pairwise linear regression of each US state's average final prevalence versus the fraction of the population in the corresponding vaccination attitude category according to surveys. This measurement is performed for the considered range of vaccination scenarios as given by the vaccination rate $\alpha$. Error bars represent the 95$\%$ C.I.}
\end{figure*}

To analyze this, we perform a pairwise correlation analysis between each state's prevalence $r(\infty)$ and the fraction of the state's population in different vaccination attitudes across the extensive range of vaccination scenarios examined earlier in this section (Figure \ref{fig:main_vaccination_curves}), as determined by the vaccination rate $\alpha$. Figure \ref{fig:main_correlation_curves} shows Pearson's correlation coefficient $\rho$ for each vaccination category as a function of $\alpha$. The main continuous lines represent the $\rho$ values, while the error bars denote the $95\%$ confidence intervals. Although the errors are relatively large due to the small number of data points (equal to the number of states, plus the District of Columbia, and the US level), the overall trends remain clear.

For the low to middle range of vaccination scenarios considered, $\alpha\in[10^{-4},6.5\times 10^{-3})$, the factor (vaccination category) that achieves greater explanatory power in terms of the Pearson coefficient is the `already' category. That is, the initial condition $v(0)$ for the vaccinated population. In terms of our framework, they populate the initial condition $n_A(0)$ (together with those in the `soon' category). Strictly speaking, though, this is not a behavioral attitude towards the vaccine that plays a role in the implemented opinion dynamics, but one of the categories in which individuals were classified when responding to the surveys in \cite{lazer2021covid}. However, a varying value of $v(0)$ across states encapsulates heterogeneous predispositions of individuals in those regions as well as different vaccination efforts deployed in each state. As vaccination rates further increase $\alpha$, a crossover appears between categories `already' and `soon', taking the `already' category the dominant role in terms of explanatory power for a higher range of $\alpha$ values explored. When vaccination rates are high enough, states with a higher proportion in this vaccination category will see their vaccination coverage greatly increased due to this behavioral category for which vaccination is just a spontaneous process, since they are initially convinced independently of their social peers. People in the `someone' category, on the other hand, would react to the vaccination process as if it were a simple contagion process: they just need an individual to change their attitude from hesitant to convinced and take the vaccine when available. For the most part, the contribution to explanatory power is very poor, especially in the lower $\alpha$ range. This is understandable for two reasons: i) according to the data (see Figure \ref{fig:supp_threshold} in Appendix \ref{app:attitudes}), this is the lowest populated vaccination category across the majority of states, and ii) these individuals require only one vaccinated neighbor to be convinced to vaccinate, thus they do not constitute a concerning bottleneck to trigger behavioral cascades. 

The next vaccination attitude in terms of increasing hesitancy is `most'. Individuals in this category are those who genuinely make their decision based on a complex contagion process for which $\theta>0.5$. As we saw in the first part of this work when studying the system's behavior under a homogeneous activation threshold $\theta$, a value of $\theta>0.5$ can have an important epidemiological impact in the population unless the initial support for the vaccine $n_A(0)$ is sufficiently high, also conditioned to the vaccination effort $\alpha$. According to the results in Figure \ref{fig:main_correlation_curves}, the impact of `most' stays fairly constant, with the Pearson coefficient $\rho$ around $0.5$ for the whole range of $\alpha$. The explanatory power is high, but other factors, either `already' for low to middle $\alpha$, or either `soon' or `never' for higher $\alpha$ show a greater relevance in these terms. What mitigates a potential negative impact of the `most' category is the fact, as it happened with `someone', that people in this category, although with a higher representation, are far from a critical fraction of the population in their respective states. Finally, the `never' category, the group of people who show the highest hesitancy or anti-vaccine sentiment, acquires the greatest relevance in the middle to the higher end of the explored range of $\alpha$. With a high enough vaccine rollout, the states that will experience higher outbreaks will be those with a higher fraction of individuals who would not take the vaccine, logically blockading an almost full percolation of the vaccinated status across the network.

Indeed, we can appreciate by looking at the data that in several states the combined fractions of the `most' and `never' attitudes amount to less than $0.5$ of the population. This indicates that in scenarios with rapid vaccination, individuals with a `most' attitude are likely to find most of their neighborhood already vaccinated, prompting their shift to a pro-vaccine stance. Second, in most states, the fraction of the population already vaccinated or willing to be vaccinated as soon as possible, $v(0)+n_A(0)$, is more than $0.4$ and often exceeds $0.5$. Including the `someone' category which, as we stated, requires just one vaccinated neighbor per individual, this vaccinated/pro-vaccine segment increases to over $0.5$ in most cases. This suggests that, from moderate $\alpha$ scenarios onwards, vaccination-opinion cascades can readily grow, leading to widespread vaccination coverage and significantly mitigated outbreaks.

\section{Discussion}
\label{sec:discussion}

In this work, we have explored the interplay of contagion processes that differ in their transmission mechanisms: a classical pairwise SIR epidemic spread; and a vaccination model driven by a social complex contagion process based on the Watts-Granovetter threshold model. Far from an arbitrary theoretical problem, the model proposed is grounded in US surveys on vaccine hesitancy, which showed complex contagion features in the attitudes of individuals towards vaccine uptake \cite{lazer2021covid}.

We have characterized the system's behavior by the relative size of the activation cascade $\Delta n_A(\infty)$, the vaccination coverage $v(\infty)$, and the disease's prevalence $r(\infty)$ at the absorbing state in the control parameter space of $n_A(0)$, initial fraction of people convinced to take the vaccine, and $\theta$, the activation threshold from hesitancy to convinced. First, under the assumption of homogeneous activation threshold $\theta$ across the population, within a realistic range of vaccination values, the emergence of a disease-free phase is possible for very low $\theta$ and medium $n_A(0)$. However, even under highly unrealistic vaccine rollout, sizeable outbreaks persist if the initial support $n_A(0)$ is not high enough and the activation threshold is moderate. In the second part, we have relaxed the homogeneity assumption by endorsing the population with heterogeneous thresholds based on real vaccine hesitancy surveys. We observed mixed behavior depending on the vaccination scenario. Generally, and as expected, individuals who are already vaccinated play a key role in reducing the epidemic's final size. Conversely, those who significantly delay vaccine uptake or directly refuse to vaccinate have a negative impact on the disease's prevalence. Fortunately, given that initial support for vaccine uptake is sufficiently high across states, the influence of anti-vaccine zealots is mitigated. 

One criticism that could be raised in our approach is the fact of resorting to an age-multilayer contact structure while the dynamical processes running on it, as here defined, do not explicitly depend on age, and moreover, without analyzing the system behavior by age groups. First, we opted to focus on global results to draw attention to the interplay of the epidemic and the threshold-based opinion dynamics rather than on a more epidemiological account of the problem. Second, we find that the main conclusions apply in more simple data-agnostic contact structures, such as homogeneous or heterogeneous single-layer networks (details in Appendices \ref{app:homogeneous_erba} and \ref{app:heterogeneous_erba}). However, this relates to an assumption carried in our model initialization, and it is the fact of assigning pro-vaccine stances randomly in the population. It is known that vaccine hesitancy can greatly vary by age, gender, or socioeconomic status \cite{troiano2021vaccine, morales2022gender}, thus if this is the case, interesting differences in terms of vaccine coverage and prevalence could be observed in populations with differences in their demographic pyramids. The lack of enough data granularity (vaccination attitudes fractions by state and age) has prevented us from pursuing this more detailed approach but it would be indeed a highly valuable complement for future work.

Finally, although the threshold model proposed is grounded in responses from real surveys, human decision-making is multifaceted, and additional mechanisms undoubtedly influence the choice to vaccinate. These mechanisms might include rational action, as is typically considered in vaccination game approaches \cite{wang2015coupled, iwamura2018realistic}, where individuals weigh the costs of different options. Additionally, the influence of mass media and social networks is nowadays critically significant, where figures of authority or online communities may exert more influence than peers or acquaintances \cite{yuan2017cyber, wilson2020social, jain2022dynamics, rathje2022social}. Moreover, these platforms can also facilitate the rapid spread of misinformation and fake news, further complicating the landscape of decision-making. Furthermore, communities evolve and adapt, often leading to fragmentation between opposing viewpoints and the creation of echo chambers \cite{schmidt2018polarization, jennings2021lack, muller2022echo}. These factors are increasingly pivotal in shaping attitudes toward vital issues and warrant consideration in future studies.

In summary, interesting potential extensions could be added to advance our understanding of the interplay between epidemics and social contagion processes related to vaccine uptake, but as it is, our model stands as an example of naturally competing simple and complex contagions processes, this last inspired by real-life observation of people's social behavior from vaccine hesitancy surveys. This is valuable in itself given the scarcity of real examples of contagion processes where transmission is genuinely characterized by group interactions.

\section{Methods}
\label{sec:methods}

\subsection{Epidemiological model}
\label{subsec:epidemic}

The epidemiological model used is the standard and well-known susceptible-infected-removed SIR model running on a population of individuals structured on the age-based multilayer network as explained previously. 

We consider discrete-time dynamics with unitary time step $\Delta t=1$ day. Then, at every time step, susceptible individuals can experience a simple pairwise contagion interaction and become infected with probability $\beta\Delta t$:
\begin{equation}
\mathcal{S}+\mathcal{I}\xrightarrow{\beta}\mathcal{I}+\mathcal{I}.
\end{equation}
Since every individual can interact with any other of their first neighbors in the network, the total probability of becoming infected is $P(\mathcal{S}\to\mathcal{I})=1-(1-\beta\Delta t)^{I_i(t)}$, where $I_i(t)$ denotes the number of infected neighbors of individual $i$ at time $t$, and $\beta$ is the disease's transmission rate. The quantity $I_i(t)$ is bounded by the node's degree $k_i$. 

Lastly, recovery or removal from the infection dynamics is, as usual, a spontaneous transition modeled as a Poisson process:
\begin{equation}
\mathcal{I}\xrightarrow{\mu}\mathcal{R}.
\end{equation}
Then, for every infected individual, the probability of decaying to the recovered/removed state is $P(\mathcal{I}\to\mathcal{R})=\mu\Delta t$ at every time step.

\subsection{Opinion dynamics model}
\label{subsec:opinion}

The SIR process runs simultaneously with two other intimately related dynamical processes: the vaccination campaign and the opinion dynamics on vaccination. Vaccination by itself is modeled as a spontaneous process proceeding continuously as long as there are eligible individuals to be vaccinated and without any preferential target. The catch is that in order to be eligible for vaccine uptake, the individual must be in a convinced or active state. Here is where the opinion formation process enters. 

To model the opinion dynamics we consider the Watts-Granovetter threshold model of social contagion \cite{granovetter1978threshold, watts2007influentials}. In this model, individuals can be classified into two states: inactive and active. Within this context, we will refer to the inactive ones as hesitant, being them denoted by $\mathcal{H}$, and we will retain the active nomenclature to refer to the pro-active ones regarding vaccine uptake, denoted by $\mathcal{A}$. As in the original model, the state $\mathcal{A}$ is irreversible, and therefore the dynamics are ruled by the transition:
\begin{equation}
    \mathcal{H}\xrightarrow{}\mathcal{A}.
\end{equation}
Mapping the status $\mathcal{H}$ to $0$ and $\mathcal{A}$ to $1$, and designating  as $o_i$ the individual's $i$ opinion, the updating rule in the original formulation of the model proceeds in the following way:
\begin{equation}
o_i(t+1)=
    \begin{cases}
        1 & \text{if } \sum_{j\in\Omega_i}\frac{o_j(t)}{k_i}\geq\theta_i,\\
        0 & \text{otherwise}.
    \end{cases}
\label{eq:watts}
\end{equation}
Here, the summation extends to $\Omega_i$, which is the neighborhood of individual $i$, $o_j$ is the status of the neighbor $j$, and $k_i$ is the degree or number of neighbors of $i$; finally, $\theta$ is the individual's activation threshold ($0\leq\theta_i\leq 1)$. Due to its lack of symmetry, if the focal individual status is already at $1$, nothing happens. Thus, as aforementioned, once the state $1$ is adopted, it is conserved until the end of the dynamical process.

For the present work, however, we do not follow the rule described in equation \ref{eq:watts} but a variation. Rather than adopting a proactive attitude towards vaccination based on the vaccine views of peers, people will make the decision on vaccine uptake based on their neighbors' vaccination status. Therefore, the updating proceeds as follows:
\begin{equation}
o_i(t+1)=
    \begin{cases}
        1 & \text{if } \frac{V_{\Omega_i}(t)}{k_i}\geq\theta_i,\\
        0 & \text{otherwise}.
    \end{cases}
\label{eq:watts_mod}
\end{equation}
Where $V_{\Omega_i}(t)$ is the total number of vaccinated individuals within $i$'s neighborhood. This change directly connects the success of the ongoing vaccination campaign to the opinion on vaccination, setting a co-evolving feedback loop between vaccination and opinion dynamics. 

Finally, every convinced or active susceptible $(\mathcal{S},\mathcal{A})$ will be vaccinated
\begin{equation}
    \mathcal{S\xrightarrow{\alpha} V},
\end{equation}
with probability $P(\mathcal{S}\to\mathcal{V})=\alpha\Delta t$.

\subsection{Vaccine hesitancy data}
\label{subsec:hesitancy}

In order to adapt the categorical attitudes to our numerical approach, as defined in the surveys by Lazer et al. \cite{lazer2021covid}, we translate them into quantifiable activation thresholds, denoted by $\theta$. Individuals in the ``already vaccinated" category are exempted from further opinion dynamics, effectively setting their activation threshold at zero. The same is presumed for individuals categorized as ``as soon as possible". We posit that these individuals are already convinced and merely awaiting their opportunity to be vaccinated; hence, we assign them a $\theta=0$. For those who would get vaccinated ``after at least someone I know", we interpret this as requiring one known individual to be vaccinated, leading to a threshold of $\theta=1/k_i$, where $k_i$ is the individual's number of first neighbors. Those in the ``after most people I know'' category are assigned a simple majority criterion, corresponding to $\theta=0.5$. It could be argued that this approach is somewhat lenient with the constraint, considering that the majority criterion might vary on an individual basis.  Lastly, for the ``would not get the COVID-19 vaccine" category, we assign a threshold of $\theta=1^+$, signifying that their decision remains unchanged irrespective of the vaccination status of their entire neighborhood, indicative of anti-vaccine sentiment. Table \ref{tab:thresholds} presents a summary of the proposed mappings.

\begin{table}[!ht]
\centering
\begin{tabular}{cc}
\hline
\textbf{survey category} & $\theta$ \\
\hline
\textit{already vaccinated} & $0$ \\
\textit{as soon as possible} & $0$ \\
\textit{after at least someone I know} & $1/k_i$ \\
\textit{after most people I know} & $0.5$ \\
\textit{would not get the [COVID-19] vaccine} & $1^+$ \\
\hline
\end{tabular}
\caption[Vaccination attitude categories and activation threshold attribution]{\textbf{Vaccination attitude categories and activation threshold attribution}. Mapping to inform the individuals' activation thresholds in our model based on the vaccination attitude categories as given by Lazer et al. \cite{lazer2021covid}.}
\label{tab:thresholds}
\end{table}

The specific fractions in every vaccination attitude category for every state in the US are depicted in Figure \ref{fig:supp_threshold} of the supplementary material \ref{app:attitudes}

\subsection{Simulations}
\label{subsec:simulations}

Our results are obtained through extensive discrete-time Monte Carlo simulations. For the sake of clarification, we describe the algorithm followed to implement the coupled dynamics described above:
\begin{enumerate}
    \item A multilayer network with $N$ nodes is generated according to the method briefly described above and developed in \cite{aleta2020data}. The individual with the highest degree is assumed to be the patient zero, which is the initial seed that will trigger the epidemic spreading process. An initial fraction $n_A(0)$ of pro-vaccine individuals are randomly assigned across the network. 
    \item At every time step $t$, every node $i$ in the contact multilayer network is visited, and the threshold model updating rule is applied (Eq. \ref{eq:watts_mod}). In practice, a list of hesitant/inactive and susceptible individuals is kept since they are the only type of agents that could change their opinion status. Opinion updating follows a parallel scheme.
    \item During the same time step, the transitions related to the SIR+V model occur. Susceptible and proactive individuals who were not vaccinated undertake a Bernoulli trial with each infected neighbor. Effective updating of health statuses also takes place following a parallel scheme.
    \item If the population of infected individuals drops to zero, the process terminates, otherwise, we move to the next time step $t\to t+1$ (and back to item 2).
\end{enumerate}

In all the experiments performed, each result for a given set of control parameters has been derived by averaging over $25$ network realizations and $50$ iterations of the dynamical process for each network.

\subsection{Code and data availability}
\label{subsec:code}

The code developed for the data curation process, simulations, analysis of results, and figure generation, as well as the curated data to feed the model, is hosted at \url{https://github.com/phononautomata/threshold}. 

\begin{acknowledgements}
A.A. and Y.M. were partially supported by the Government of Aragon, Spain, and ERDF "A way of making Europe" through grant E36-20R (FENOL), and by Ministerio de Ciencia e Innovación, Agencia Española de Investigaci\'on (MCIN/AEI/ 10.13039/501100011033) Grant No. PID2023-149409NB-I00. A.A. acknowledges support through the grant RYC2021-033226-I funded by MCIN/AEI/10.13039/501100011033 and the European Union ‘NextGenerationEU/PRTR’.
\end{acknowledgements}

\clearpage

\appendix
\renewcommand{\thefigure}{\Alph{section}\arabic{figure}}

\section{Fraction of every vaccine attitude per state (survey data)}
\label{app:attitudes}

\setcounter{figure}{0}

\begin{figure*}
\centering
\includegraphics[width=\textwidth]{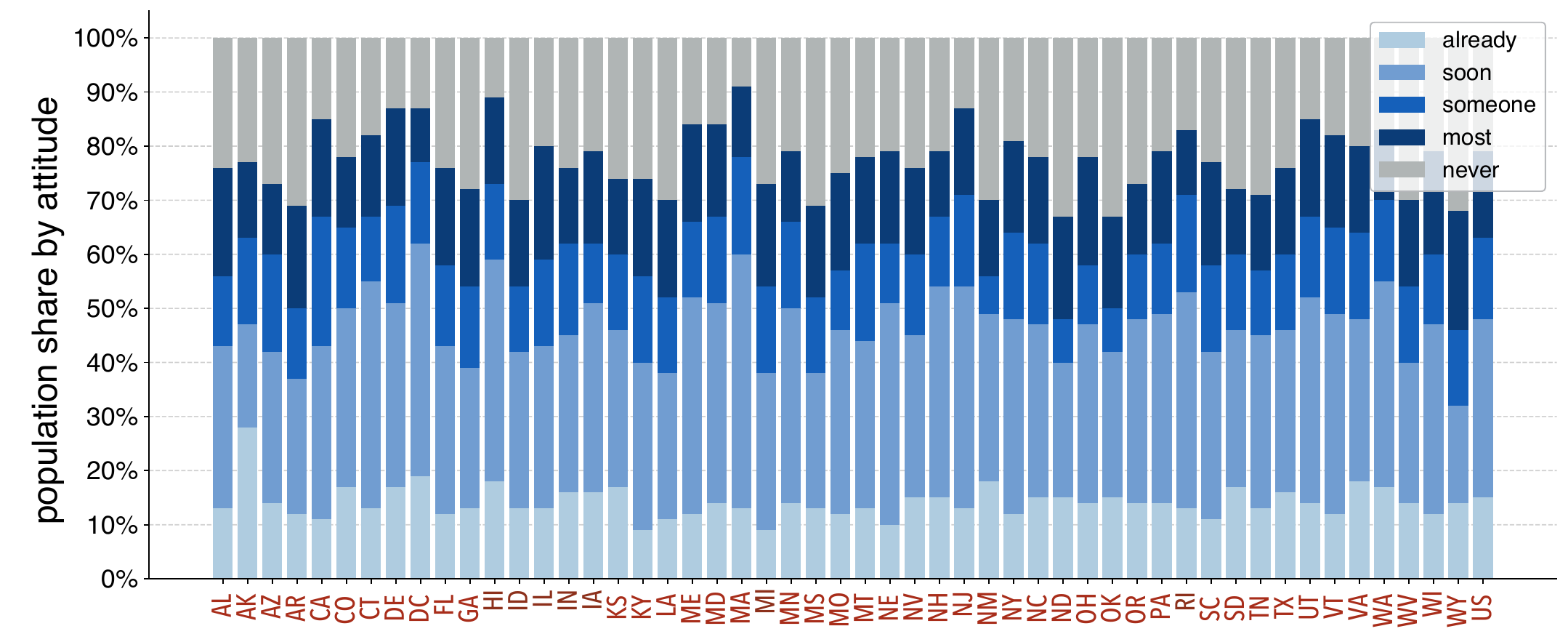}
\caption{\label{fig:supp_threshold} \textbf{US states' population fractions by attitude towards vaccination}. From the bottom to the top, for each state the bar segments represent the following attitudes towards vaccination: ``already vaccinated'', ``as soon as possible'', ``after at least some people I know'', ``after most people I know'', ``I would never take the [COVID-19] vaccine'', as obtained from Lazer et al. \cite{lazer2021covid}.}
\end{figure*}

In figure \ref{fig:supp_threshold}, we report the fraction of the population with each vaccination attitude according to the survey by Lazer et al.\cite{lazer2021covid}. 

In Figure \ref{fig:supp_threshold}, we report the fraction of the population in each vaccination attitude category according to the survey by Lazer et al.\cite{lazer2021covid}. This survey provides valuable insights into public sentiment regarding vaccination across various states in the United States. The data are broken down into five distinct attitudes toward vaccination:
\begin{itemize}
    \item ``Already vaccinated": This category includes individuals who have already received the COVID-19 vaccine.
    
    \item ``As soon as possible'': These individuals are ready to get vaccinated as soon as they are eligible, reflecting a proactive approach toward vaccination.
    
    \item ``After at least some people I know'': These individuals are hesitant but open to getting vaccinated once they observe others around them, such as friends or family, receiving the vaccine.
    
    \item ``After most people I know'': This group is even more hesitant and prefers to wait until a larger portion of their social circle gets vaccinated.

    \item ``I would never take the [COVID-19] vaccine'': The most resistant group, comprising individuals who are strongly opposed to taking the COVID-19 vaccine under any circumstances.
\end{itemize}

\section{Complementary results for homogeneous thresholds}
\label{app:sections}

\setcounter{figure}{0}

\begin{figure*}
\centering
\includegraphics[width=\textwidth]{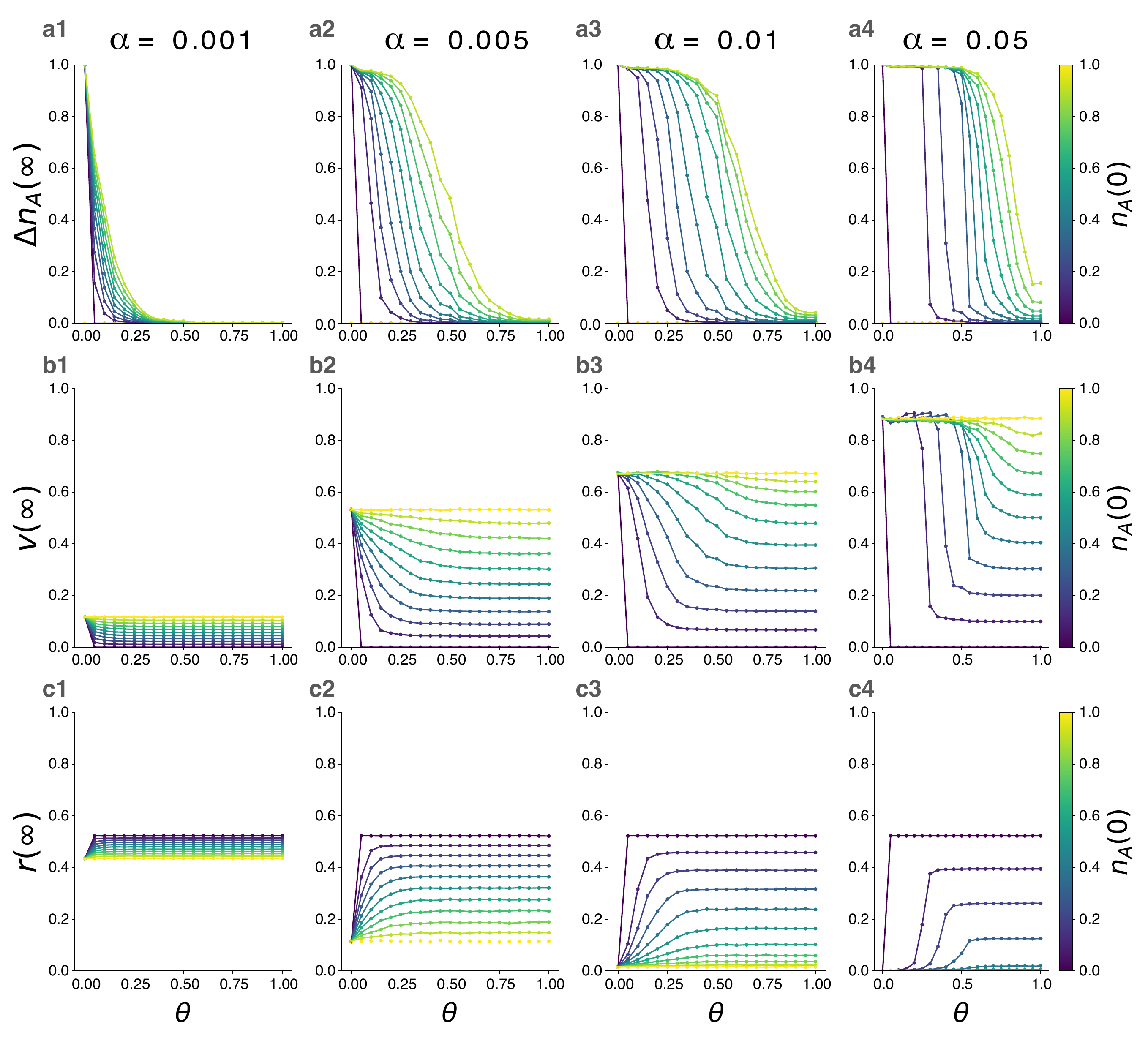}
\caption{\label{fig:supp_sections} \textbf{Sections for the US multilayer network under homogeneous thresholds}. Panels 1 to 4 represent different vaccination efforts $(\alpha=0.001$, $0.005$, $0.01$, $0.05$ from left to right columns). Panels \textbf{a} depict $\Delta n_A(\infty)$, panels \textbf{b} represent $v(\infty)$, and $\textbf{c}$, $r(\infty)$ curves for constant $n_A(0)$ (value given in the color bar) and varying $\theta$.}
\end{figure*}

Complementarily to the heatmap shown in Figure \ref{fig:main_homogeneous_heatmap}, we represent in Figure \ref{fig:supp_sections} the system's behavior for the US multilayer network as sections of constant $n_A(0)$ for varying $\theta$ under chosen vaccination scenarios.

Now the first row (Panels a1-a4) represents the evolution of $\Delta n_A(\infty)$ for the proposed vaccination scenarios in the main text. As $\alpha$ increases, we observe how $\Delta n_A(\infty)$ increases too, being the section with higher $n_A(0)$ those that sustain higher $\Delta n_A(\infty)$ values for a larger range of increasing $\theta$. Whereas the lowest vaccination scenario cannot trigger a substantial behavioral cascade, since all the curves quickly decay to $0$ for low $\theta$, higher $\alpha$ provokes successful behavioral cascades as confirmed by looking at the evolution of $v(\infty)$ curves in the second row (Panels b1-b4). The greatest change among the considered scenarios is observed when increasing effort from $\alpha=0.001$ to $\alpha=0.005$, with $v(\infty)\approx 0.1$ changing to $v(\infty)>0.5$, for the extreme case of $n_A(0)=1$. Even for sections with a high level of initial hesitancy (low $n_A(0)$), the change in $v(\infty)$ is relevant. Further increases of $\alpha$ naturally lead to higher $v(\infty)$, but the gains are clearly sublinear. As for the disease's prevalence (Panels c1 to c4), we observe, as reported in the main text, a very high (and narrow) impact for the lowest vaccination scenario, and as we increase $\alpha$, we appreciate the relevance of a different $n_A(0)$, and the fact that a disease-free phase is reached when this initial support $n_A(0)$ is low enough (clearly the transition occurs between $\alpha=0.005$ and $\alpha=0.1$). 

Finally, these sections allow us to clearly appreciate the abrupt nature of the changes in $\Delta n_A(\infty)$, $v(\infty)$ and $r(\infty)$ for varying $\theta$. Interestingly, as $\alpha$ increases, changes are more abrupt, as the competition between the different dynamic processes involved intensifies.

\section{Complementary results for surveys' heterogeneous thresholds}
\label{app:scatters}

\setcounter{figure}{0}

\begin{figure*}
\centering
\includegraphics[width=\textwidth]{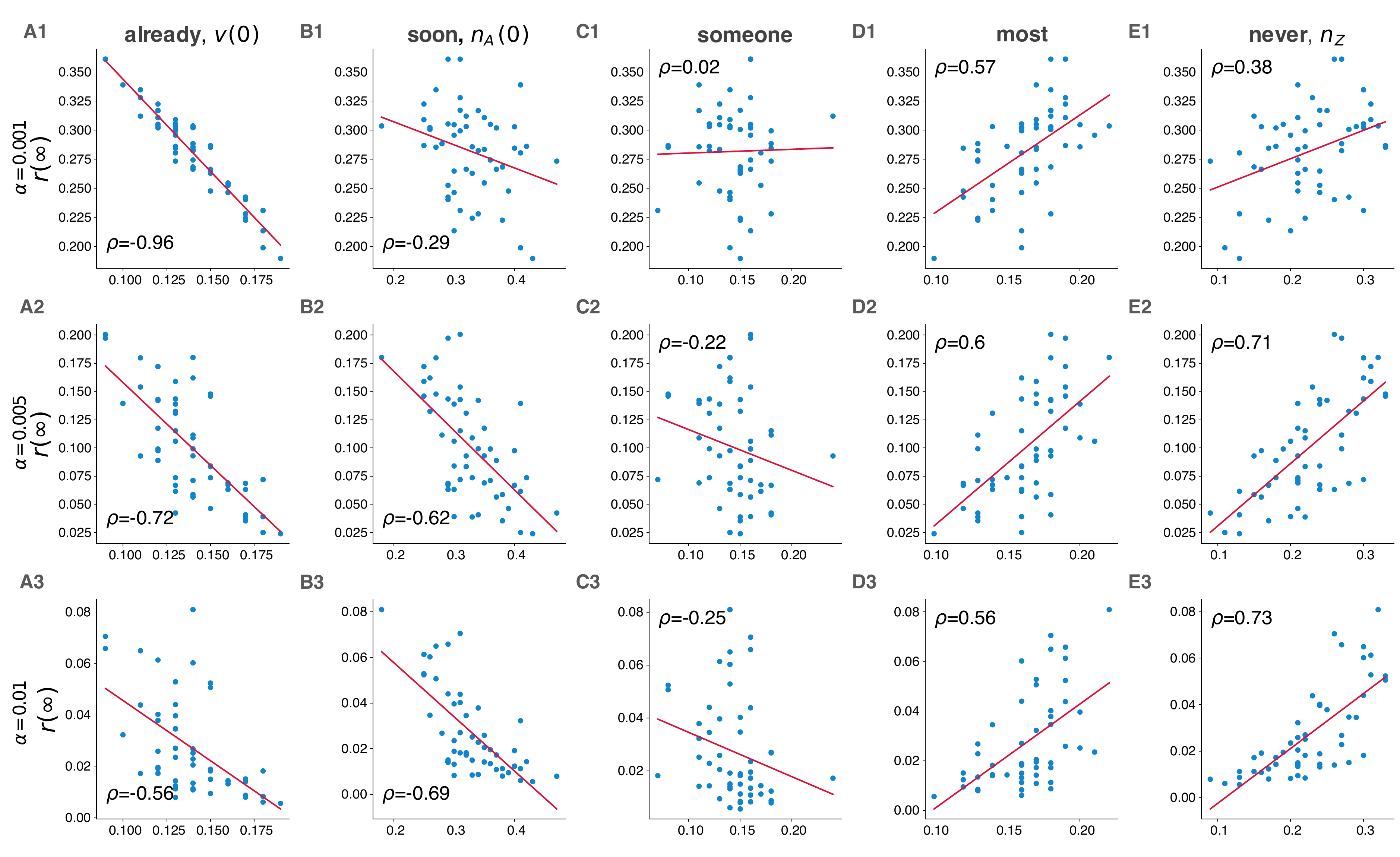}
\caption[Pairwise scatter plots of prevalence and vaccine attitudes]{\label{fig:supp_multilayer_scatters}\textbf{Pairwise scatter plots of prevalence and vaccine attitudes}. Panels with letters A, B, C, D, and E show, respectively, the scatter plots of prevalence (y-axis) against vaccine attitudes (x-axis): ``already vaccinated'', ``as soon as possible'', ``after at least some people I know'', ``after most people I know''. ``I would never take the vaccine''. The state of Alaska has been removed as an outlier due to the markedly lower epidemiological impact due to the fact that its high initial condition $v(0)$.}
\end{figure*}

Figure \ref{fig:supp_multilayer_scatters} depicts the specific scatter plots of prevalence versus population fraction in vaccination attitude categories for three selected vaccination scenarios ($\alpha=0.001$, $0.005$, and $0.01$).

\clearpage

\section{Homogeneous thresholds on simple Erd\H{o}s-Renyi and Barabási-Albert networks}
\label{app:homogeneous_erba}

\setcounter{figure}{0}

\begin{figure}[!ht]
\centering
\includegraphics[width=1.0\linewidth]{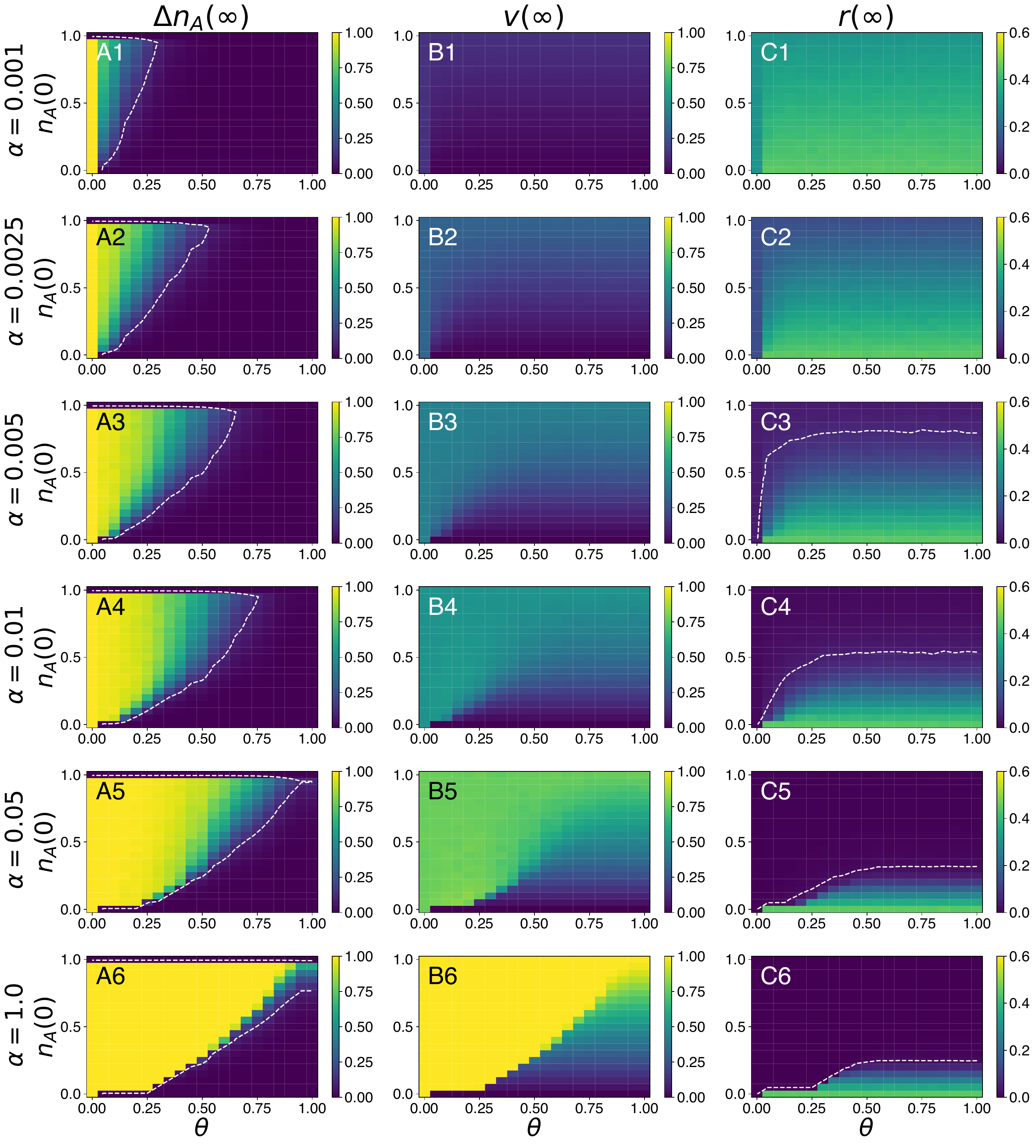}
\caption[Homogeneous thresholds in Erd\H{o}s-Rényi networks]{\textbf{Homogeneous thresholds in Erd\H{o}s-Rényi networks}. Results for normalized activation change $\Delta n_A(\infty)$, vaccination coverage $v(\infty)$, and prevalence $r(\infty)$. In every panel, outcomes are explored in $(\theta,n_A(0))$ space. Every row shows the solution for a different vaccination rate $\alpha$. Every point in the diagrams amounts to $25$ network realizations times $25$ dynamical realizations. White dashed line contours mark an approximate (due to finite size effects) separation between $\Delta n_A(\infty)=0$ or $r(\infty)=0$ and non-null values of the respective observables.}
\label{fig:supp_homo_er}
\end{figure}

\begin{figure}[!ht]
\centering
\includegraphics[width=1.0\linewidth]{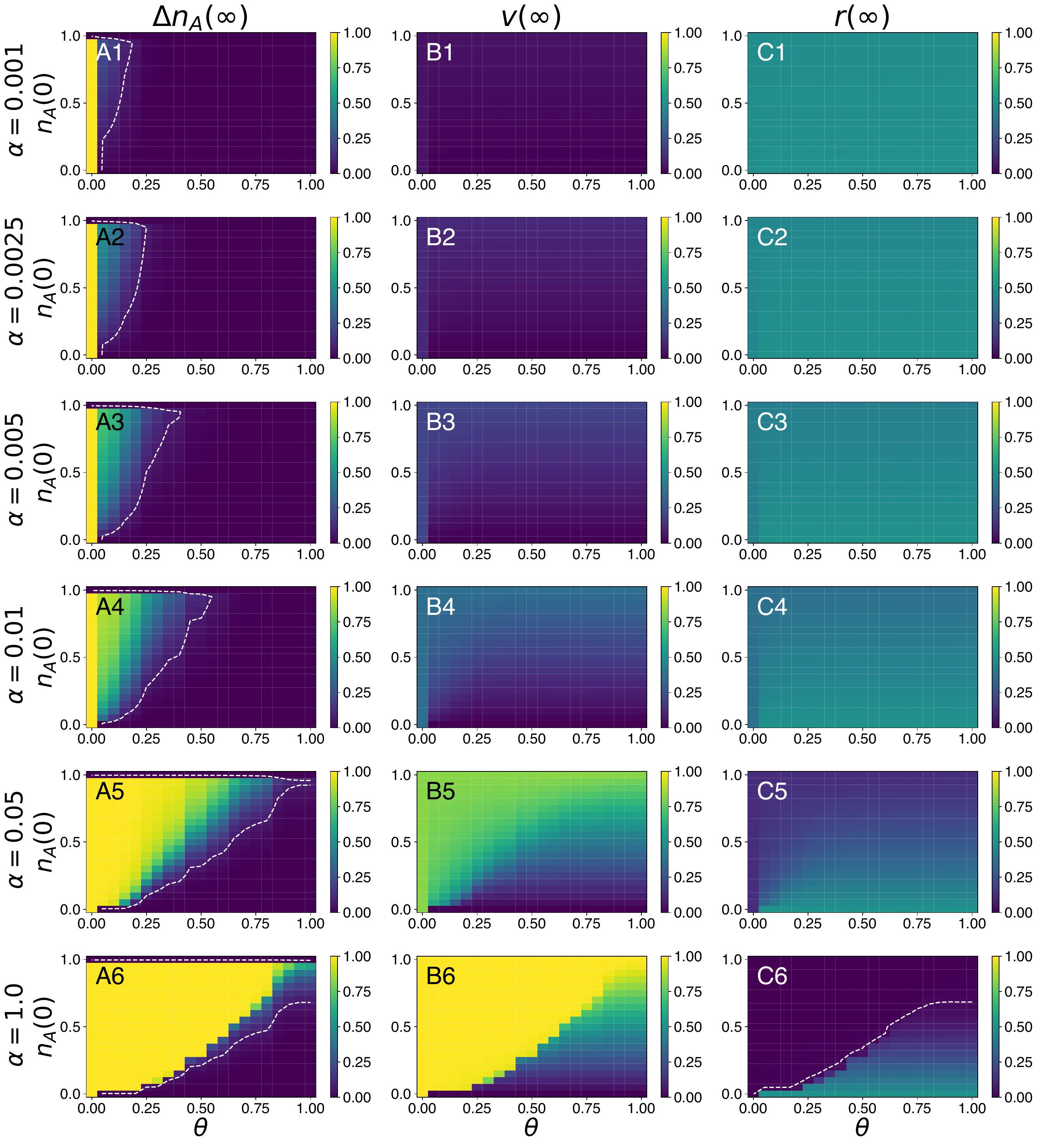}
\caption[Homogeneous thresholds in Barabási-Albert networks]{\textbf{Homogeneous thresholds in Barabási-Albert networks}. Results for normalized activation change $\Delta n_A(\infty)$, vaccination coverage $v(\infty)$, and prevalence $r(\infty)$. In every panel, outcomes are explored in $(\theta,n_A(0))$ space. Every row shows the solution for a different vaccination rate $\alpha$. Every point in the diagrams amounts to $25$ network realizations times $25$ dynamical realizations. White dashed line contours mark an approximate (due to finite size effects) separation between $\Delta n_A(\infty)=0$ or $r(\infty)=0$ and non-null values of the respective observables.}
\label{fig:supp_homo_ba}
\end{figure}

As a complement to our main text results, we also show results of the same coupled dynamics under a homogeneous activation threshold $\theta$ in Erd\H{o}s-Rényi and Barabási-Albert networks both with average degree $\langle k\rangle=10$.

\subsection{Erd\H{o}s-Rényi networks}
\label{subapp:homogeneous_erdosrenyi}

Although not presented here, the trivial scenario with $\alpha=0$ triggers no cascading behavior, and subsequently, no vaccination at all (since $v(0)=0$). Therefore the prevalence is always the maximum expected according to the given disease and network characteristics. In this case, for the stochastic simulations on the ER network, we obtain an average value $r_{ER}^{\mathrm{max}}\equiv r_{ER}(\infty)=0.454$, with $[0.450, 0.460]$ as the $95\%$ confidence interval (CI). Indeed, for the limit case $n_A(0)\to 0$ and $\theta>0$, with $\theta>n_A(0)$, vaccination is trivially null for any $\alpha$ and the system's prevalence approaches the case with $\alpha=0$, $r_{ER}(\infty)\to r_{ER}^{\mathrm{max}}(\infty)$. As a further reference, given the values being used as epidemiological parameters, $\beta=0.03$ for the transmission rate and $\mu=0.2$ for the removal rate, and the social interaction term $\langle k\rangle=10$, under a homogeneous mixing approximation we would expect $R_0=\beta\langle k\rangle/\mu=1.5$, and thus $r_{hom}(\infty)=0.58$ as computed from the classical SIR solution given by the transcendental equation $r(\infty)=1-e^{-R_0r(\infty)}$. Despite ER networks being regarded as homogeneous systems that can resemble more a standard well-mixing approach than, for instance, scale-free networks where the degree distribution has a power-law form, the relatively low $\langle k\rangle$ results in a significant contact saturation that leads to a reduced epidemiological impact\footnote{It can be shown that as $\langle k\rangle$ grows (while preserving the product $\beta\langle k\rangle$), the simulation outcomes approach the homogeneous mixing assumption, and then $r_{ER}(\infty)\to r_{hom}(\infty)$.}.

For the lowest proposed vaccination rate scenario, $\alpha=0.001$, we observe that for the most of the $(\theta, n_A(0))$-space, there is no growing support for vaccine adoption, that is, $\Delta n_A(\infty)=0$ (Figure \ref{fig:supp_homo_er} panel A1). We refer to the space where $\Delta n_A(\infty)=0$ to the \textit{impassive} region. Consequently, vaccination coverage is extremely low (Figure \ref{fig:supp_homo_er} panel B1), and sizeable outbreaks emerge, ranging roughly between $r(\infty)\in(0.3, 0.46)$ (Figure \ref{fig:supp_homo_er} panel C1). The region of the parameter space where adherence to vaccine uptake is the maximum possible, $\theta=0$ with $n_A(0)\in(0,1)$, shows a slightly larger vaccination coverage and a lower prevalence. In any case, all the parameter space is characterized by an endemic solution ($r(\infty)\neq 0$).

As the vaccination rate $\alpha$ increases, there is a growing region of $(\theta,n_A(0))$, where hesitancy recedes and fades away, and consequently vaccine uptake emerges and the disease prevalence decreases. For $\alpha=0.0025$, the maximum VC attained is at $v(\infty)=0.324$ $[0.318, 0.327]$ (Figure \ref{fig:supp_homo_er} panel B2), and the corresponding prevalence is reduced to $r(\infty)=0.145$ $[0.140, 0.149]$ (Figure \ref{fig:supp_homo_er} panel C2). The impact has been notably reduced in some regions of the parameter space, but the vaccination rates are still slow enough to avoid the emergence of epidemic outbreaks. Doubling $\alpha$, $\alpha=0.05$, clearly, the solution landscape changes. We attain a vaccination coverage around $v(\infty)=0.425$ $[0.417, 0.432]$ (Figure \ref{fig:supp_homo_er} panel B3), which provokes the beginning of an emergent disease-free region (Figure \ref{fig:supp_homo_er} panel C3). Still increasing $\alpha$ values translates into successful positive feedback between the vaccination adoption opinion dynamics and the vaccination campaign, which in turn increases the parameter space region where a disease-free solution reigns and, consequently, the endemic solution is bounded to domains with high adoption threshold $\theta$ and low to very low initial support $n_A(0)$ (Figure \ref{fig:supp_homo_er} panels C4, C5, and C6, progressively). 

Overall, for a fixed value of $\theta$, moving from $n_A(0)=0$ to $n_A(0)=1$, means that hesitancy loses ground, vaccination coverage increases, and disease prevalence tends to zero. In high enough $\alpha$ scenarios, for low $\theta$, the transitions occur in a more abrupt way, whereas for medium to high $\theta$, the transitions are smoother. Varying $\theta$ with fixed $n_A(0)$ does not have a notable effect except where a phase separation exists. If the initial vaccine acceptance $n_A(0)$ is high enough, even $\theta\to 1$ has no effect on vaccination and prevalence. A giant component of vaccinated individuals can emerge fast enough to avoid a sizeable outbreak. On the other hand, as $n_A(0)$ decreases (for a fixed $\alpha$), critical values of $\theta$ appear that, if surpassed, can bring the system from a disease-free phase to an endemic phase. This critical threshold $\theta$, however, can be pushed toward higher values if the vaccination campaign proceeds at faster rates.

The extreme and unrealistic case of $\alpha=1$, depicts an abrupt transition when looking at the behavior of $\Delta n_A(\infty)$ in $(\theta, n_A(0))$-space (Figure \ref{fig:supp_homo_er} panel A6). A rather marked boundary separates the region where the maximum size of cascading behavior occurs and the region where factoring in the initial support $n_A(0)$, there is no further change induced. This \textit{impassive} region precludes the system from reaching total VC there. However, for the largest part of it (roughly when $n_A(0)>0.25$), VC, as being propelled by a high vaccination rate, is large enough to drive the system to the free-disease phase. Finally, it is noteworthy that the solution landscape for $r(\infty)$ is very similar across the vaccination scenarios with $\alpha=0.05$ and $\alpha=1$ (Figure \ref{fig:supp_homo_er} panels C5 and C6, respectively). This observation suggests the existence of diminishing returns to the vaccination process as determined by $\alpha$, possibly constrained by the underlying network topology. Therefore, a moderate vaccination may be sufficient to efficiently mitigate the epidemic impact of the disease. 

\subsection{Barabási-Albert networks}
\label{subapp:homogeneous_barabasialbert}

For $\alpha=0.001$, the situation is qualitatively similar to the coupled dynamics running on homogeneous networks. Vaccination rates are too low to trigger a cascading behavior and change individuals' hesitant opinions toward vaccine-proactive individuals. Virtually, VC is null across the $(\theta, n_A(0))$ (Figure \ref{fig:supp_homo_ba} panel B1), and prevalence is maximal for the BA network under the current epidemiological parameters, being $r_{BA}(\infty)=0.504$ $[0.503, 0.505]$ (Figure \ref{fig:supp_homo_ba} panel C1). However, at odds with the ER network case, continuing to increase $\alpha$ has hardly an effect. Adherence to vaccination advances very slowly (Figure \ref{fig:supp_homo_ba} panels A1 to A4), and consequently VC is sluggish. Even at $\alpha=0.01$, disease prevalence continues to be very high across all the $(\theta, n_A(0))$ space. Still, even at $\alpha=0.05$, an important region of the parameter space suffers a high epidemic impact, and no disease-free solution emerges. Lastly, for $\alpha=1$, the situation of $\Delta n_A(\infty)$ (Figure \ref{fig:supp_homo_ba} panel A6) and $v(\infty)$ (panel B6) now resembles more closely the results for the ER network (Figure \ref{fig:supp_homo_er} panels A6 and B6, respectively). However, in the impassive region ($\Delta n_A(\infty)=0$) where VC remains very low to null, the epidemic impact is larger than in homogeneous networks.

In BA networks, the competition between disease contagion and the opinion-vaccination dynamics is overwhelmingly won by the former. This outcome is expected, considering the well-documented fact that the topology of heterogeneous networks facilitates a rapid spread of disease \cite{moreno2002epidemic}. In this coupled dynamical system under study, disease transmission is a simple contagion process, requiring only a single contact between an infected and a susceptible individual to propagate, whereas opinion dynamics—a complex contagion—demands a critical mass of individuals who are not only persuaded but also vaccinated, to initiate a cascading effect of behavioral change. In contrast to the dynamics on ER networks, the heavy-tailed degree distribution found in BA networks means that highly connected nodes play a significant role and necessitate a larger number of influenced neighbors to shift their stance toward a pro-vaccination viewpoint, thereby impeding or delaying widespread vaccination coverage.

\section{Survey-based thresholds on ER and BA networks}
\label{app:heterogeneous_erba}

\setcounter{figure}{0}

\subsection{System's behavior under different vaccination scenarios}
\label{subapp:heterogeneous_erba_impact}

To finalize the analysis on ER and BA networks, we extend the model to the survey-based thresholds scenario as we did in the main text for the data-driven multilayer.

\begin{figure}[!ht]
\centering
\includegraphics[width=1.0\linewidth]{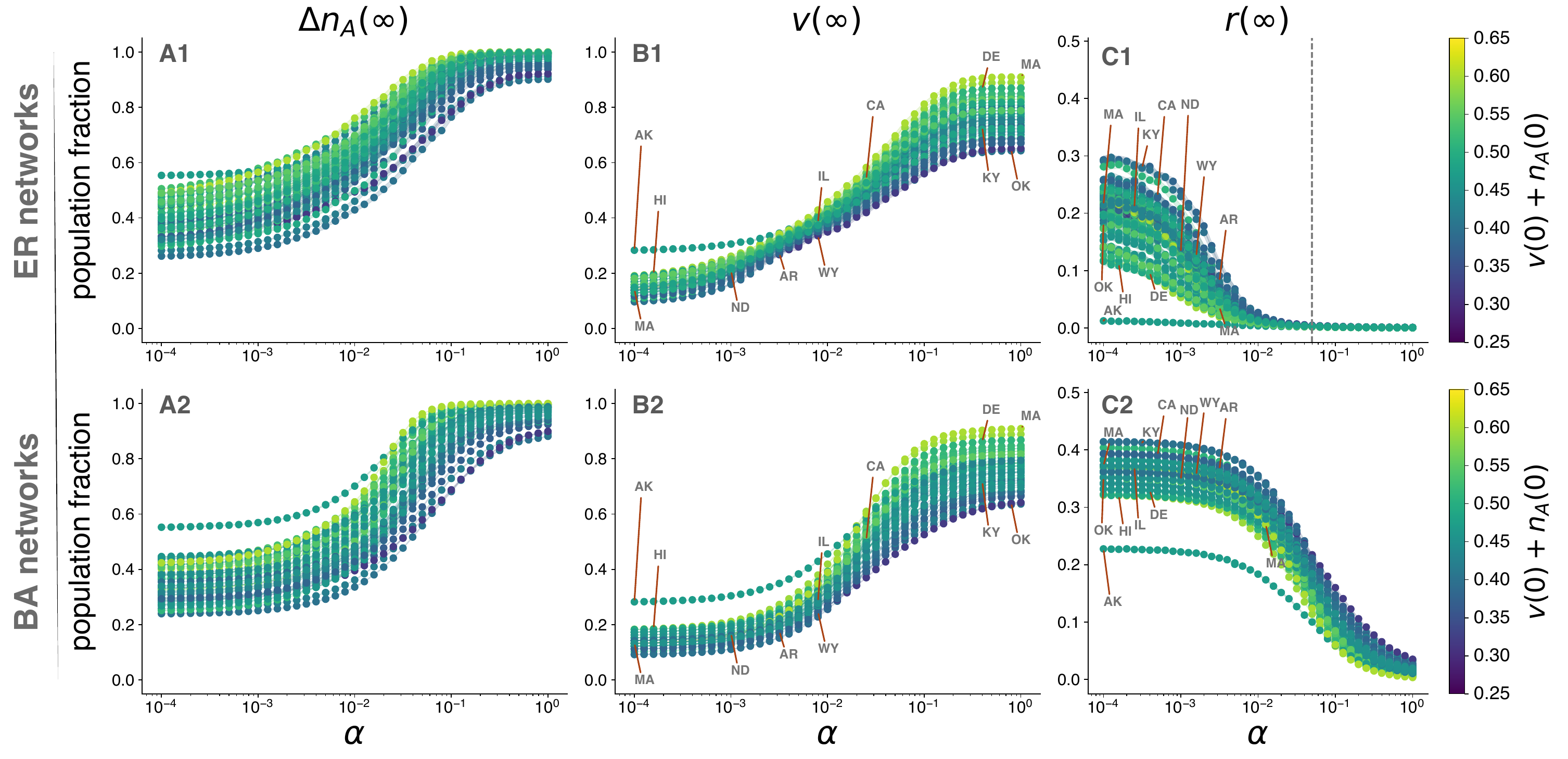}
\caption[Epidemiological impact across US states for varying vaccination rates $\alpha$]{\textbf{Epidemiological impact across US states for varying vaccination rates $\alpha$}. Top panels: Results across US states on ER networks. The vertical dashed line represents an approximate separation between the endemic phase and the disease-free phase. Bottom panels: BA networks. Panels A1 and A2: $\Delta n_A(\infty)$ vs. $\alpha$. Panels B1 and B2: $v(\infty)$ vs. $\alpha$. Panels C1 and C2: $r(\infty)$ vs. $\alpha$. The color bar represents the initial fraction of support for the vaccine: ``already vaccinated'' $v(0)$ plus ``as soon as possible'' $n_A(0)$. Some state labels are shown for reference.}
\label{fig:supp_vaccination}
\end{figure}

In Panels B1 and B2 of Figure \ref{fig:supp_vaccination}, we focus on vaccination coverage. In low $\alpha$ regimes, VC is primarily between $0.1$ and $0.2$, with the exception of Alaska (AK). For intermediate $\alpha$ values, representing more realistic vaccine uptake rates, the disparity in VC across states reduces but widens again as $\alpha\to 1$. Due to inherent differences in each state's $\langle k\rangle$ and, more importantly, their $\theta$ distribution, the curves intersect at various points. These varying gaps and curve crossings for different $\alpha$ values reflect the diversity in vaccine attitudes across states. As a specific example, the case of Massachusetts (MA) is interesting. In the lowest $\alpha$ regimes, MA occupies a median rank among states, with $13\%$ of its population having already received the vaccine. However, with increasing $\alpha$, the vaccination-opinion feedback loop is further stimulated. This facilitates the progressive vaccination of individuals in subsequent categories, specifically ``soon'' at $47\%$, and ``someone'' at $18\%$. Ultimately, this leads MA to emerge as one of the states with the highest vaccination coverage.

Regarding the final disease prevalence $r(\infty)$, Panels C1 and C2 display results for ER and BA networks, respectively. The most notable difference between the network types aligns with previous analyses. For ER networks (Figure \ref{fig:supp_vaccination} C1), in very low $\alpha$ regimes, prevalence across US states ranges from $0.1$ to $0.3$. These values decrease rapidly with increasing vaccination deployment, reaching a disease-free state beyond $\alpha\approx 0.05$. Alaska is correspondingly an exception here too; its prevalence is negligible across all vaccination rate scenarios, as represented by $\alpha$. With an initial ``already vaccinated'' fraction of $v(0)=0.29$, Alaska's outcomes suggest that for very low $\alpha$, the opinion-vaccination cascades do not initiate, resulting in $v(\infty)\approx v(0)$. This indicates that such an initial $v(0)$ is close to the herd immunity threshold $p_{c}$, preventing sizeable outbreaks. For a well-mixed population, akin to an ER network with a basic reproduction number $R_0=1.5$, the herd immunity threshold is $p_c=1-1/R_0\approx 0.33$. Alaska's negligible impact is thus understandable in light of its initial condition $v(0)$\footnote{It should be noted that $v(0)$ represents an initial condition in a population otherwise naive to the disease. In reality, however, the ``already vaccinated'' category in surveys preceded the second wave of COVID-19 in the US.}. Conversely, in BA networks we saw the impossibility of attaining a disease-free solution for most $\alpha$ values. Now, even Alaska experiences significant outbreaks for most $\alpha$ values. In the case of Massachusetts, previously highlighted for its vaccination coverage (VC) progression with varying $\alpha$ in both ER and BA networks, a similar trend is evident in its prevalence data. It begins at a mid-rank position but becomes one of the states least impacted when $\alpha$ reaches high (yet realistic) values.

Overall, while the relationship is not strictly monotonic, it is evident that larger $v(0)+n_A(0)$ values correspond to higher VC and lower prevalence, and vice versa, underscoring the significance of initial vaccine support conditions in influencing the epidemiological impact.

\subsection{Relevance of heterogeneous thresholds}
\label{subapp:heterogeneous_erba_relevance}

\begin{figure*}
\centering
\includegraphics[width=\textwidth]{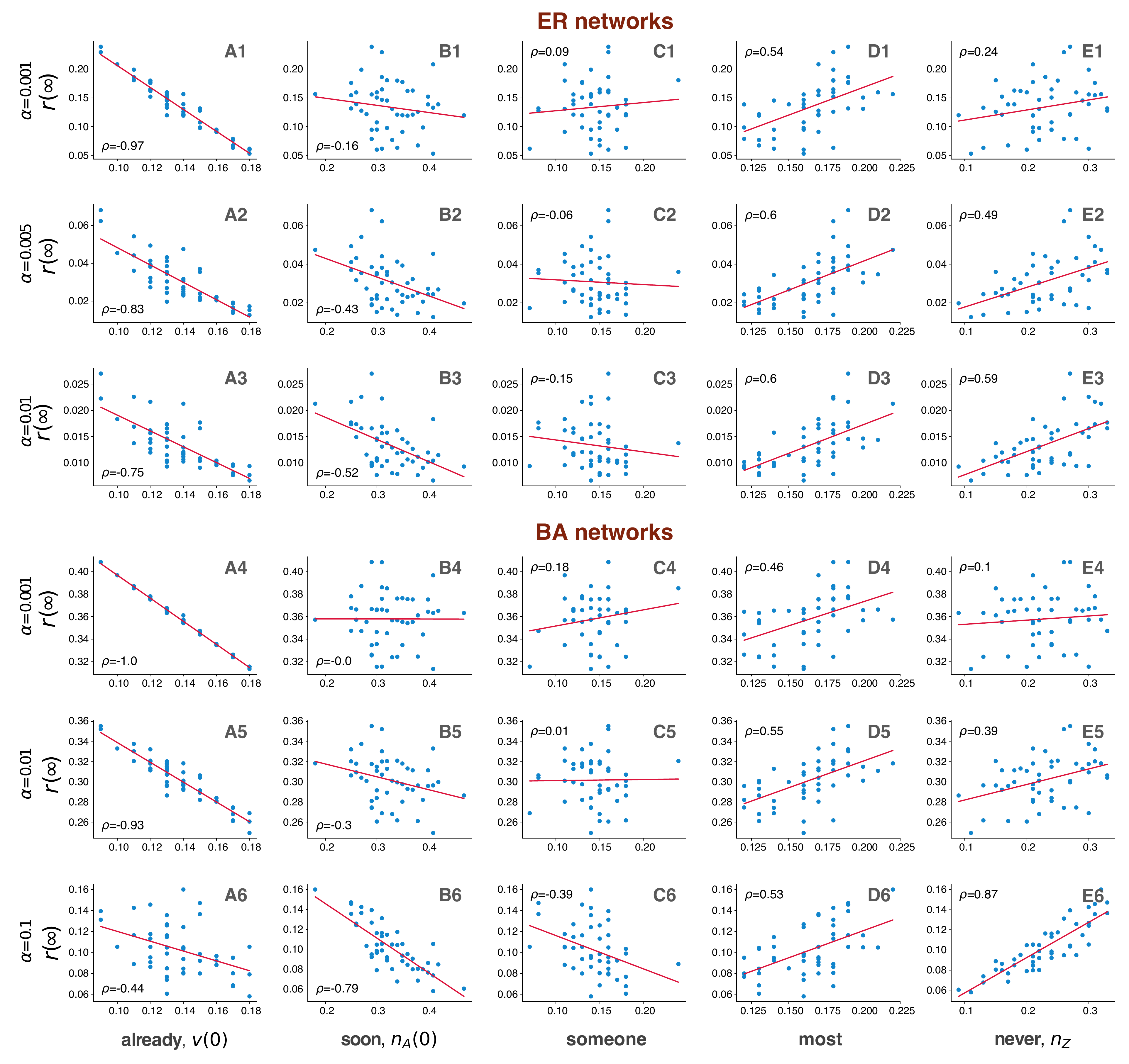}
\caption[Pairwise scatter plots of prevalence and vaccine attitudes]{
\label{fig:supp_erba_scatters}
\textbf{Pairwise scatter plots of prevalence and vaccine attitudes}. Panels with letters A, B, C, D, and E show, respectively, the scatter plots of prevalence (y-axis) against vaccine attitudes (x-axis): ``already vaccinated'', ``as soon as possible'', ``after at least some people I know'', ``after most people I know''. ``I would never take the vaccine''.
}
\end{figure*}

We explicitly show scatter plots for the different vaccination attitude categories and each US state's final epidemic size for selected vaccination scenarios (see Figure \ref{fig:supp_erba_scatters}).

We observe a clear correlation as indicated by Pearson's correlation coefficient for low and moderate $\alpha$ scenarios of $r(\infty)$ and $v(0)$ (``already vaccinated'') (refer to Figure \ref{fig:supp_erba_scatters} A1 to A6) and a relatively small or very poor correlation with respect to the ``soon'' category (Figure \ref{fig:supp_erba_scatters} B1 to B6), the only exception being $\alpha=0.1$ for the BA network case (Panels A6 and B6). As expected, the higher $v(0)$ is, the lower $r(\infty)$. Particularly in the lowest vaccination rate scenarios, the correlation between $r(\infty)$ and $v(0)$ is very high (Panel A1 with $\rho=-0.97$) or even absolute (Panel A4 with $\rho=-1$), in contrast to $r(\infty)$ vs. $n_A(0)$. In these scenarios, where the vaccination campaign is irrelevant, it is evident that the emphasis is on the already vaccinated portion of the population rather than those pro-vaccine but not yet vaccinated. We observe that as $\alpha$ increases, the Pearson coefficient's explanatory power for $v(0)$ diminishes, while it grows (in absolute value) for $n_A(0)$.

For the attitude ``someone'' (I will take the vaccine after someone I know), we observe mixed results. At the lowest $\alpha$, it indeed correlates positively with prevalence, and as $\alpha$ increases, the sign of the correlation changes, and $\rho$ grows in magnitude. However, correlations are generally weak, suggesting that these individuals are not key in shaping the spread. 

The ``most'' attitude (I will take the vaccine after most people I know) shows moderate but distinct correlations. In states where this fraction of individuals is larger, larger outbreaks tend to occur. Across different $\alpha$ scenarios and networks tested, the correlation coefficient $\rho$ varies within a small range. One reason these individuals are not more influential is that if a cascade of pro-vaccine individuals grows quickly enough, those with the ``most'' attitude will also likely become pro-vaccine, thus helping to reduce the final epidemic size. On the other hand, the presence of such individuals would be highly detrimental if it were to be accompanied by either high anti-vaccine sentiment or low pro-vaccine attitudes.

Finally, we consider the anti-vaccine attitude population fraction (``never''), i.e. the zealots, denoted by $n_Z$. A higher $n_Z$ correlates with a higher $r(\infty)$, although their statistical correlation varies depending on the vaccination scenario. Similar to other attitudes, the fraction of zealots or anti-vaccine individuals increases its explanatory power under more intensive vaccination campaigns. Indeed, when vaccination becomes less relevant, it is logical to expect that individuals refusing vaccines under any circumstances will not have a significant impact. Their influence becomes particularly notable in BA networks for high to very high vaccination rates.

Although not directly comparable, assuming a uniform threshold setting in ER networks with initial vaccination support between $0.4$ and $0.6$, we see in Figure \ref{fig:supp_homo_er} (Panel C5) that we are within the disease-free phase of the diagram. Similarly, Figure \ref{fig:supp_vaccination} (Panel C1) shows that all states reach a disease-free solution at approximately $\alpha=0.05$. This suggests that the initial conditions across states are favorable enough to effectively yield reduced prevalence, provided vaccination progresses swiftly. Conversely, for BA networks, consistently with the simplified homogeneous setting previously explored, even such robust initial support is insufficient to significantly lower the system's prevalence. 

\clearpage

\bibliography{references}

\end{document}